\documentclass[aps,pra,reprint,groupedaddress,showpacs]{revtex4-1}

\usepackage[T1]{fontenc}
\usepackage{graphicx}
\usepackage[utf8x]{inputenc}
\usepackage{amssymb,amsmath}
\usepackage{siunitx}
\usepackage[overload]{empheq}
\usepackage{subfigure}
\usepackage{numprint}
\usepackage{bm}
\usepackage{tikz}
\usetikzlibrary{%
      decorations.pathreplacing,%
      decorations.pathmorphing,%
  shapes%
  }
\usepackage{pgfplots}
\usepackage[active]{srcltx}  
\usepackage{tabularx}
\usepackage{xcolor} 
\definecolor{halfgray}{gray}{0.55} 
\definecolor{webgreen}{rgb}{0,.5,0}
\definecolor{webbrown}{rgb}{.6,0,0}
\definecolor{lgreen} {RGB}{180,210,100}
\definecolor{dblue}  {RGB}{20,66,129}
\definecolor{lred}   {RGB}{220,0,0}
\definecolor{nred}   {RGB}{224,0,0}
\definecolor{norange}{RGB}{210,129,34}
\definecolor{nyellow}{RGB}{255,221,0}
\definecolor{ngreen} {RGB}{98,158,31}
\definecolor{dgreen} {RGB}{78,138,21}
\definecolor{nblue}  {RGB}{28,130,185}
\definecolor{jblue}  {RGB}{20,50,100}

\newcommand{\e}{\mathrm{e}}
\newcommand{\ic}{i}
\newcommand{\B}{\boldsymbol}
\newcommand{\tens}[1]{\B{#1}}
\newcommand{\re}{\mathfrak{Re}}
\newcommand{\im}{\mathfrak{Im}}
\newcommand*\conj[1]{
  \vbox{
  \hrule height 0.3pt
  \kern0.5ex
  \hbox{
  \kern-0.4em
  \ifmmode#1\else\ensuremath{#1}\fi
   \kern-0.em
  }
 } 
}

\newcommand{\grad}{\bm{\mathrm{\nabla}}}
\renewcommand{\div}{\bm{\mathrm{\nabla\cdotp}}}
\newcommand{\Mop}{\mathcal{M}}
\newcommand{\Lop}{\mathcal{L}}
\newcommand{\ddroit}{\mathrm{d}}
\newcommand{\source}{\mathcal{S}}
\newcommand{\bra}{\ensuremath{\left\langle}}
\newcommand{\ket}{\ensuremath{\right\rangle}}
\newcommand{\mb}{\ensuremath{\,\middle| \,}}
\newcommand\abs[1]{\left|#1\right|}

\begin{document}

\title{Quasimodal expansion of electromagnetic fields in open two-dimensionnal structures}

\author{Benjamin Vial}
\email[]{benjamin.vial@fresnel.fr}

\author{Frédéric Zolla}

\author{André Nicolet}

\author{Mireille Commandré}

\affiliation{Aix Marseille Universit\'e, CNRS, Centrale Marseille, Institut Fresnel, UMR 7249, 13013 Marseille, France}

\date{\today}

\begin{abstract}
A quasimodal expansion method (QMEM) is developed to model and understand the scattering properties of arbitrary shaped 
two-dimensional (2-D) open structures.
 In contrast with the bounded case which have only discrete spectrum (real in the lossless media case), 
 open resonators show a continuous spectrum 
 composed of radiation modes and may also be characterized by resonances associated to complex eigenvalues (quasimodes). 
 The use of a complex change of coordinates to build Perfectly Matched Layers (PMLs) allows the numerical computation of 
 those quasimodes and of approximate radiation modes. Unfortunately, the transformed operator at stake is no longer 
self-adjoint, and classical modal expansion fails. To cope with this issue, we consider an adjoint
eigenvalue problem which eigenvectors are bi-orthogonal to the eigenvectors of the initial problem. 
The scattered field is expanded on this complete set of modes leading to a reduced order model of 
the initial problem. The different contributions of the eigenmodes to the scattered field unambiguously appears through the modal coefficients,
allowing us to analyze how a given mode is excited when changing incidence parameters. 
This gives new physical insights to the spectral properties of different open structures such as 
nanoparticles and diffraction gratings. Moreover, the QMEM proves to be extremely efficient for the computation 
of Local Density Of States (LDOS).
\end{abstract}


\maketitle

\section{Introduction}

Resonance is a central phenomenon in every field of wave physics and is related to what is commonly called a spectral problem 
(the eigenfrequencies and eigenmodes solutions of source free governing equations). 
These spectral elements can be understood as privileged vibrational states and are thus an intrinsic characteristic of
the system. Closed cavities with perfect conducting walls have real eigenvalues and normal modes, 
but for open electromagnetic systems, even for materials without losses, eigenfrequencies $\omega$ are in general complex, 
the real part $\omega'>0$ giving the resonant frequency and the imaginary part $\omega''<0$ the linewidth of the resonance. 
The associated leaky modes \cite{Sammut76} (also known as resonant states \cite{fox1961resonant,simon_resonances}, quasimodes \cite{lamb1964theory},
quasi-normal modes \cite{Settimi2009,Settimi2003}, 
quasi-guided modes \cite{Tikhodeev2002} in the literature) are proportional to $\cos[\omega'(t-r/v)]\exp[\omega''(t-r/v)]$ so they 
are no longer of finite energy and even grow exponentially in space at infinity while possessing finite lifetime. 
Physically, this exponential divergence corresponds to a wavefront excited at past times 
and propagating away from the system, and the infinite energy can be understood as the accumulation of the energy radiated from the open resonator to 
the rest of the universe.\\

The study of resonant properties of open optical systems is of fundamental interest in various domains of application 
such as biophotonics \cite{Rigneault2005,Wenger2008} for single molecule fluorescence detection, 
antennas \cite{Muehlschlegel2005,Rolly2012}, photonic crystals \cite{Yoshie2001,Akahane2003}, 
microstructured optical fibers \cite{Nicolet2}, diffraction gratings \cite{Peng1996,Tamir1997,Bonod2008} and 
subwavelength aperture arrays \cite{RevModPhys.82.729,Genet1}
for example for filtering 
applications \cite{FEHREMBACH2003,fehrembach, Ding2004a,these_vial_2013_anglais}, quantum electrodynamics 
(QED) cavity experiments \cite{Gleyzes2007,Kuhr2007,CavityQED,Vuckovic2001}, \textit{etc..}. Finding eigenmodes 
of open structures with non trivial geometries is thus of great theoretical and practical interest.\\

It is well known that the eigenfrequencies of an open system correspond to the poles of its scattering matrix 
or of Fresnel coefficients \cite{nevbook}. 
The numerical computation of these poles remains a challenging task and several approaches have been used. 
Firstly, one has to compute the S-matrix coefficients, which can be done by numerous numerical method: 
the {Rigorous Coupled
Wave Method} (RCWA \cite{gaylord,rokushima1}) also known as {Fourier
Modal Method} (FMM \cite{li3,guizal2009reformulation}), 
the Differential Method \cite{watanabe}, 
the Integral Method \cite{maystre}, the 
Finite Difference Time Domain method (FDTD \cite{yee,saj}), the 
{Finite Element Method} (FEM \cite{delort,Ohkawa1,bao,Demesy2009}), 
the {Method of Fictious Sources} (MFS \cite{zolla2})\dots 
Secondly, one must find the poles of the S-matrix, and several approaches have been developed to do so: 
computing the poles of its determinant \cite{Centeno2000}, the 
 poles of its maximum eigenvalue \cite{Felbacq2001}, others techniques based on the linearization of its inverse \cite{Gippius2010}, or more recently 
 an iterative method \cite{Bykov13}. In spite of numerous ways of improving the convergence of these methods, the dimension of the S-matrix has to be 
 very large in general to guarantee a sufficient precision of the results, which can lead to numerical instabilities.
 Note that another method based on the computation of Cauchy path integrals of S-matrix valued functions of a complex variable
 can be used to find an arbitrary number of poles in a given region of the complex plane \cite{Felbacq2001,fpcf}.\\
 
 For a given problem, one can define an associated Maxwell's operator that depends on geometry, material properties and 
 boundary conditions. We are interested here in operators associated with functional spaces with elements defined on an unbounded 
 domain. In that case, it turns out that the \emph{spectrum} of this operator (the generalized set of eigenvalues) has to be 
 considered to fully characterize the resonant properties of the problem at stake. Particularly, in
 addition with quasimodes associated with discrete complex eigenfrequencies, the spectrum of such an operator shows a real continuous part
 associated with radiation modes expressing the propagation of energy from the structure towards the infinite space. \\
 
 We use a finite element spectral method to study the resonant properties of open optical systems. Thanks to 
 its versatility it can handle complex geometries and arbitrary materials, which is necessary in most practical applications. 
Moreover, the method naturally leads to a linear eigenvalue problem in matrix form after discretization because the basis functions 
are frequency independent, in contrast to other methods such as the Boundary Element Method (BEM) 
where the equations are projected on frequency dependent Green functions. 
 The FEM has already been used to compute leaky modes in different cases \cite{Nicolet2,Eliseev2005,Zhang2008}, 
 however, it is of prime importance to use adequate absorbing boundary conditions to correctly handle 
 the divergent behaviors of fields. The solution is to use Perfectly Matched Layers (PML \cite{Berenger1994185}) 
damping the fields in free space \cite{pmlCompel,Popovic2003,Bon-Gou-Haz-Pri-2009}. Through an \textit{ad hoc} complex change of coordinates, 
PMLs provides the suitable non-Hermitian extension of Maxwell's operator that makes possible the computation of leaky and radiation modes. 
It is worth noting that the geometrical transformation introduced to define PMLs is virtually exact and its effect is not only to turn 
the continuous spectrum into complex values but also to allow the computation of complex frequencies associated with quasimodes. 
The continuous spectrum is finally approximated by a discrete set of eigenvalues because of the discretization of the problem by the FEM 
and the effect of the truncation of PMLs at a finite distance.\\

Once the eigenmodes of the open system have been found, one expects a resonant behavior of the diffracted field 
when shining light with frequency close to the real part of a given eigenfrequency. In other words, the electromagnetic spectrum 
shows rapid variations with incident parameters (frequency and angle) around the resonant frequency, the rate of variation being related 
to the imaginary part of the eigenfrequency, accounting for the leakage of the mode. This crucial information is at 
the heart of the diffractive properties of open resonators. An interesting question is how to recover a diffracted field 
with the modes as building blocks. This can be done by expanding any diffracted field on the complete basis of the eigenmodes. \\
The question of the spectral representation of waves
in open systems have extensively been studied \cite{Shevchenko1971,Leung90,Leung98} but is still not fully addressed for the general case (with non trivial geometries and material properties),
thus making quasimodal expansion techniques not
well suited for practical applications. More recently, an approach similar (by the use of PML to treat an approximated closed problem) 
to the one reported here have been proposed \cite{Derudder1999,Derudder2000,Dai2012}. 
Another method called Resonant State Expansion (RSE \cite{Muljarov2010,Doost2012,Doost2013}) consists in treating the system as a perturbation 
of a canonical problem which spectral elements are known in closed form. 
The idea is to compute these perturbed modes and to use them in the modal decomposition. Finally, a recent approach based on quasi-normal modes expansion have been developed to define mode volumes and revisit the Purcell factor in nanophotonic resonators \cite{Lalanne2013}.\\

The major difficulty relies in the fact that the modes in open systems cannot be normalized in a standard fashion by integrating their square modulus.
 Instead we must consider an adjoint eigenproblem with Hermitian conjugate material properties, the modes of which modes are bi-orthogonal to 
 the modes of the initial problem. Equipped with this set of modes, the spectral representation of any diffracted field can be obtained. 
 The coefficients in the expansion express the coupling between the sources (particularly a plane wave) and a given mode, 
 revealing the conditions of excitation of this mode when varying incident parameters. With this QuasiModal Expansion Method (QMEM), 
 we obtain a reduced order model with a few modes that can accurately describe the diffractive behavior of open structures. 
 In addition, the source point case makes the computation of Green functions and LDOS straightforward once the eigenmodes of the systems have been found.\\
 
 The paper is organized as follows: we first expose our FEM formulation of the diffraction of a plane wave 
 by an arbitrary number of scatterers of possibly complex shape buried in a multilayer stack for both fundamentals 
 polarizations. The materials can be inhomogeneous, dispersive and anisotropic 
 and the formulation can handle mono-periodic gratings. We detail the equivalent radiation problem, 
 the use of PML and the computational parameters related to the FEM. 
 In Section \ref{sec:specpb}, we develop the formulation of the spectral problem, with emphasis 
 on the structure of the spectrum of Maxwell's operator and its modifications with the use of PML. 
 The Section \ref{sec:mm} is devoted to the set up of the QMEM through the treatment of an adjoint spectral problem. 
 Finally, we give examples of application in Section \ref{sec:exnum} showing the strength of the methods developed 
 by providing a meticulous modal analysis of scattering properties of open resonators. 
 We first study  a triangular rod in vacuum and show how the angle dependent excitation of resonances in the absorption cross section 
 can be explained by the QMEM coefficients. The modal reconstruction of diffracted field, absorption cross section and LDOS are also provided. 
 The second example is that of a lamellar diffraction grating, for which the transmission and reflection coefficients 
 show a complex spectral behavior that is fully explained and faithfully reproduced by the QMEM.

\section{Scattering problem}
\label{part:scatt_pb}
\subsection{Setup of the problem}

\begin{figure}[htbp!]
\includegraphics[width=\linewidth]{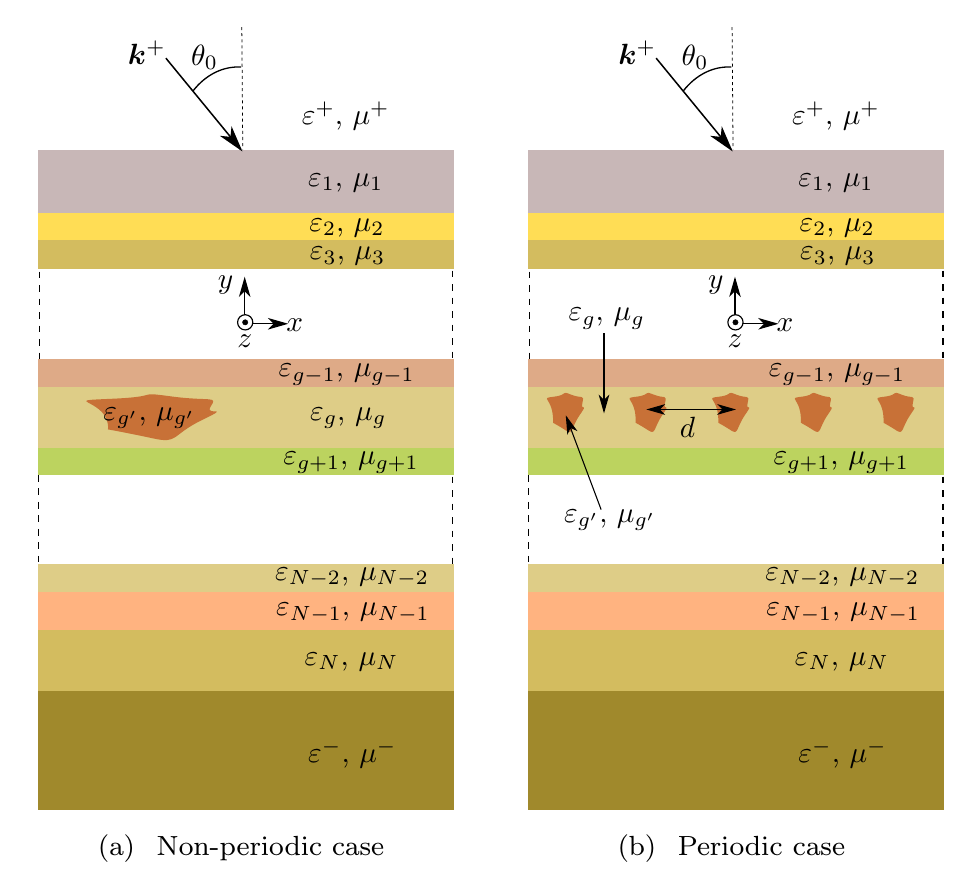}
\caption{Sketch of the studied structures and notations.}
\label{dessin}
\end{figure}

The formulation used here is the one described in Refs. \onlinecite{Demesy2007,Demesy2010}. 
It relies on the fact that the diffraction problem can be rigorously treated as an equivalent radiation problem with sources 
inside the diffractive object. We denote by $\B x$, $\B y$ and $\B z$ the unit vectors of an orthogonal Cartesian co-ordinate system $Oxyz$.
We deal with time-harmonic fields, so that the electric and magnetic fields are represented by complex
vector fields $\B E$ and $\B H$ with a time-dependence in $\mathrm{exp}(-\ic\omega t)$,
which will be dropped in the notation in the sequel. Moreover, we denote $k_0=\omega/c$.\\
To remain as general as possible (in particular to handle PML), we may consider $z$-anisotropic material, so the tensor 
fields of relative permittivity $\tens{\varepsilon}$ and
 relative permeability $\tens{\mu}$ are of the following form:

\begin{equation}
\tens{\varepsilon}=
 \left(
  \begin{array}{ c c c}
\varepsilon_{xx} & \conj{\varepsilon_{a}} & 0 \\
 \varepsilon_{a} & \varepsilon_{yy} & 0 \\
  0 &  0 & \varepsilon_{zz}
  \end{array} \right)
\text{ and}\hspace{20pt}
\tens{\mu}=
 \left(
  \begin{array}{ c c c}
\mu_{xx} & \conj{\mu_{a}} & 0 \\
 \mu_{a} & \mu_{yy} & 0 \\
  0 &  0 & \mu_{zz}
  \end{array} \right),
\end{equation}
where the coefficients $\varepsilon_{xx}$, $\varepsilon_{aa}$,...$\mu_{zz}$ are (possibly) complex valued
 functions of $x$ and $y$, and where $\conj{\varepsilon_{a}}$ (resp. $\conj{\mu_{a}}$) 
 is the complex conjugate of
$\varepsilon_{a}$ (resp. ${\mu_{a}}$).\\
The studied structures are invariant along $Oz$. They are composed of $N$ homogeneous layers 
of relative permittivity $\varepsilon_{j}$ and relative permeability $\mu_{j}$, $j=1,\dots,N$ (See Fig.~\ref{dessin}). 
These layers may contain one or several inhomogeneities. For the sake of clarity, we only consider one scatterer (See Fig.~\ref{dessin}(a)) of
or one infinitely $d$-periodic chain of scatterers (See Fig.~\ref{dessin}(b)) of isotropic and homogeneous material with 
relative permittivity $\varepsilon_{g'}$ and relative permeability $\mu_{g'}$. These restrictions are assumed to simplify the theoretical developments but 
our methods can treat additional diffractive objects buried inside different layers possibly made of $z$-anisotropic materials
without increasing the computational cost. The substrate (-) and superstrate (+) are homogeneous an isotropic with relative permittivity 
$\varepsilon^{-}$ and $\varepsilon^{+}$ and relative permeability $\mu^{-}$ and $\mu^{+}$. 
The structure is illuminated by an incident plane wave of wave vector defined by the angle $\theta_0$:
$\B k^+ =\alpha\B x+\beta\B y =k^+(\sin{\theta_0}\B x-\cos{\theta_0}\B y)$.
 Its electric (resp. magnetic) field is linearly polarized along the $z$-axis, this is the
 so-called transverse electric or s-polarization case (resp. transverse magnetic or p-polarization case). 

Under the aforementioned assumptions, the diffraction problem in a non conical
mounting can be separated in two fundamental scalar cases TE and TM.
Thus we search for a $z$-linearly polarized electric (resp. magnetic)
 field $\B E=e(x,y)\B z$ (resp. $\B H=h(x,y)\B z$). Denoting $\widetilde{\tens{\varepsilon}}$
 and $\widetilde{\tens{\mu}}$
the $2\times2$ matrices extracted from $\tens{\varepsilon}$ and $\tens{\mu}$:

\begin{equation}
 \widetilde{\tens{\varepsilon}}=
 \left(
  \begin{array}{ c c}
\varepsilon_{xx} & \conj{\varepsilon_{a}} \\
 \varepsilon_{a} & \varepsilon_{yy}
  \end{array} \right)
\hspace{10pt}\text{ and}\hspace{10pt}
 \widetilde{\tens{\mu}}=
 \left(
  \begin{array}{ c c}
\mu_{xx} & \conj{\mu_{a}}\\
 \mu_{a} & \mu_{yy}
  \end{array} \right),
\end{equation}

the functions $e$ and $h$ are solution of similar differential equations:

\begin{equation}
\Lop_{\tens{\xi},\chi}(u):=\div(\tens{\xi}\, \grad u)+k_0^2\, \chi\, u=0,
\label{helm2D}
\end{equation}
 such that $  u^d:=u-u_0$ satisfies an Outgoing Wave Condition (OWC), with
\begin{equation*}
 u=e, \hspace{5pt} \tens{\xi}=\widetilde{\tens{\mu}}^{\rm T}/\mathrm{det}(\widetilde{\tens{\mu}}),
\hspace{5pt} \chi=\varepsilon_{zz} \quad\text{for the TE case,}
\end{equation*}
\begin{equation*}
 u=h, \hspace{5pt} \tens{\xi}=\widetilde{\tens{\varepsilon}}^{\rm T}/\mathrm{det}(\widetilde{\tens{\varepsilon}}),
\hspace{5pt} \chi=\mu_{zz} \quad\text{for the TM case.}
\end{equation*}

Under this form, the problem is not adapted to a resolution by a numerical method 
because of infinite issues: the sources of the plane wave are infinitely far, the geometric domain is unbounded and 
in the periodic case the scattering structure is itself infinite. 
To circumvent these issues, we compute only the 
diffracted field solution of an equivalent radiation problem with sources inside the scatterers, 
we use PMLs to truncate the unbounded domain at
a finite distance,
and we use quasiperiodicity conditions to model a single period in the grating case.

\subsection{Equivalent radiation problem}
Denoting $\tens{\xi_1}$ and $\chi_1$ the tensor field and the scalar function describing the
multilayer problem,
the function $u_1$ is defined as the unique solution of:
\begin{equation}
 \Lop_{\tens{\xi_1},\chi_1}(u_1)=0,
\end{equation}
such that $u_1^d:=u_1-u_0$ satisfies an OWC. The expression of this function can be 
calculated with a matrix transfer formalism. The unknown function $u_2^d$ is thus given by:
\begin{equation}
 u_2^d=u-u_1=u^d-u_1^d.
\end{equation}
The scattering problem (\ref{helm2D}) can be rewritten as:
\begin{equation}
 \Lop_{\tens{\xi},\chi}(u_2^d)=-\Lop_{\tens{\xi},\chi}(u_1):=\source_1.
\label{equ:u2d}
\end{equation}
The term on the right hand side can be seen as a source term $\source_1$ with support in 
the diffractive objects and is known in closed form (See Appendix~\ref{appendix:source_term} for the detailed expression).

\subsection{Perfectly Matched Layers}

Transformation optics has recently unified various techniques in
computational electromagnetics such as the treatment of open problems, helicoidal geometries 
or the design of invisibility cloaks \cite{Nicolet2008}.
These apparently different problems share the same concept of geometrical transformation, 
leading to equivalent material properties \cite{Lassas2001739,Lassas2001}.
 A very simple and practical rule can be set up \cite{fpcf}: when changing the co-ordinate system, 
 all you have to do is to replace the initial materials properties $\tens{\varepsilon}$
 and $\tens{\mu}$ by equivalent material properties $\tens{\varepsilon}^\mathrm{s}$ and $\tens{\mu}^\mathrm{s}$ 
 given by the following rule:
\begin{equation}
\tens{\varepsilon}^\mathrm{s}=\B J^{-1}\tens{\varepsilon} \B J^{-\rm T}\mathrm{det}(\B J) \hspace{5pt}  
\text{and~} \hspace{5pt} \tens{\mu}^\mathrm{s}=\B J^{-1}\tens{\mu} \B J^{-\rm T}\mathrm{det}(\B J),
\label{equ_tranform}
\end{equation}
where $\B J$ is the Jacobian matrix of the co-ordinate transformation consisting of the partial derivatives 
of the new coordinates with respect to the original ones ($\B J^{-\rm T}$ is
the transposed of its inverse). In this framework, the most natural way to define PMLs is to consider 
them as maps on a complex space $\Gamma$, which co-ordinate change leads to equivalent permittivity
and permeability tensors. The associated complex valued change of coordinates is given by:
\begin{equation}
\eta'(\eta)=\int_0^{\eta} s_\eta(\ell)\rm d \ell,
\label{ys}
\end{equation}
where $\eta′$ is a complex coordinate such that $\re(\eta′) = \eta$
is the original coordinate (corresponding to the initial
“physical” coordinate system). The function $s_\eta$ is a complex
valued function depending on a real variable. In practice, the change
of coordinates is chosen to be the identity in the region
of interest (where the fields have therefore directly their
untransformed values) and the complex stretch is limited to
a surrounding layer. In this paper we use cylindrical or Cartesian PML with constant
stretching coefficient $s_\eta=\sigma\e^{\ic\phi}$ with $\sigma>0$ and $0<\phi<\pi/2$.\\

\subsection{Quasiperiodicity}

Let $\Gamma_l$ and $\Gamma_r$ be the two parallels boundaries orthogonal to the direction of periodicity $x$ and separated by $d$. 
Bloch theorem implies:
\begin{equation}
 u_2^d(x+d)=u_2^d(x)\e^{\ic\alpha d}.
\end{equation}
In practice, we consider $u_2^d$ as unknown on $\Gamma_l$ (which is done by applying Dirichlet homogeneous conditions) 
and we impose the value of one point on $\Gamma_d$ to be equal to the value of the corresponding point on $\Gamma_l$ multiplied by the dephasing $\e^{\ic\alpha d}$.

\subsection{The FEM formulation}
\label{subsection:eigpb}

 The radiation problem defined by
Eq.~(\ref{equ:u2d}) is then solved by the FEM, using PMLs to truncate the infinite regions and by
setting convenient boundary conditions on the outermost limits of the domain, depending on the 
problem. For mono-periodic structures, we apply Bloch quasi periodicity conditions with coefficient $\alpha$ on the two
parallel boundaries orthogonal to the grating direction of periodicity. In all cases, we apply homogeneous 
Neumann or Dirichlet boundary conditions on the outward boundary of the PMLs. 
The computational cell is meshed using $2\textsuperscript{nd}$ order Lagrange elements. In the numerical
examples in the sequel, the maximum element size is set to 
$\lambda / (N_m\sqrt{\abs{\re(\varepsilon)}})$, where $N_m$ is an integer 
(between 6 and 10 is usually a good choice). 
The final algebraic system is solved using a direct solver (PARDISO \cite{Schenk2004475}).

\section{Spectral problem}\label{sec:specpb}

\begin{figure}[htbp!]
\includegraphics[width=0.9\columnwidth]{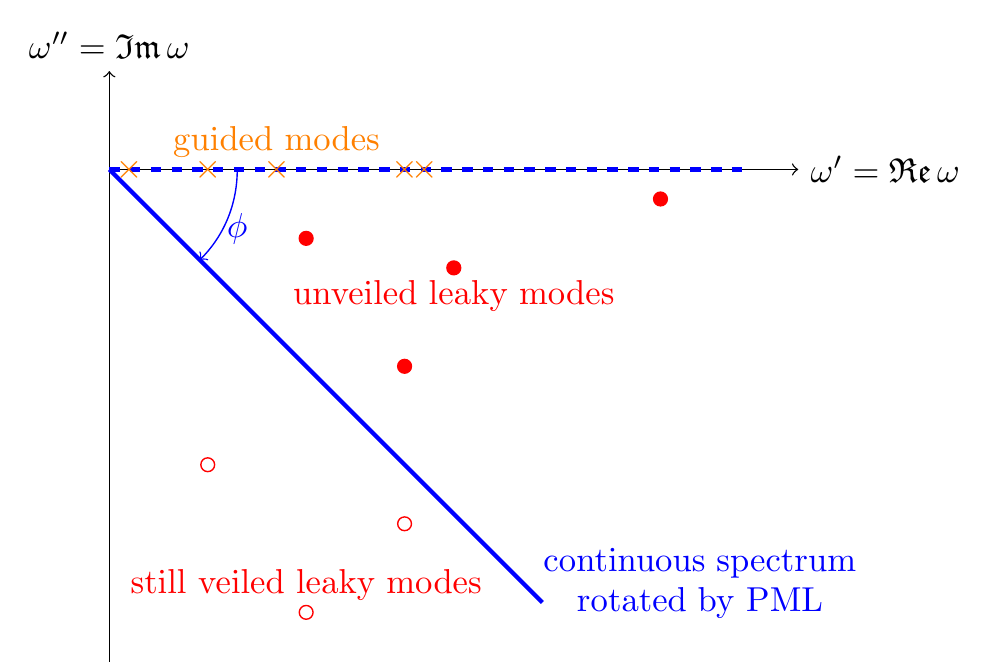}
\caption{Guided modes, continuous spectrum and leaky modes in an open waveguide.\label{PML_leaky}}
\end{figure}

Generally speaking, the diffractive properties of open systems can be studied at a more fundamental level by 
looking for both the generalized eigenfunctions and eigenvalues of a Maxwell's operator $\Mop_{\tens{\xi}}$ associated with the problem. 
The definition and classification of the spectrum of an operator is quite a delicate mathematical question and
is out of the scope of this paper (nevertheless we give 
in Appendix \ref{appendix:spec_anal} some basic definitions). 

The eigenproblem we are dealing with consists in finding the solutions of source free Maxwell's equations, \textit{i.e.}
 finding eigenvalues ${\Lambda_n}=({\omega}_n/c)^2$ and non zero eigenvectors $v_n$ such that:
\begin{equation}
\Mop_{\tens{\xi}}(v_n):=-\div(\tens{\xi}\,\grad v_n)=\Lambda_n \,\chi \, v_n,
\label{eq:eigenpb}
\end{equation}
where $v_n$ satisfies an O.W.C.
We consider here non dispersive materials, so that the eigenvalue problem (\ref{eq:eigenpb}) is linear. 
Note that in the periodic case, we search for Bloch-Floquet eigenmodes so 
the operator is parametrized by the real quasiperiodicity coefficient $\alpha$.\\

For bounded problems with lossless and reciprocal materials (with permittivity and permeability
tensors represented by Hermitian operators),
the operator $\Mop_{\tens{\xi}}$ is self-adjoint
so its eigenvalues are real, positive and discrete. For Hermitian open problems, the spectrum of 
the associated operator is \emph{real}\footnote{Note that when dealing with passive lossy materials, this spectrum moves in the lower complex plane $\omega''<0$, 
but if active materials are considered, the eigenfrequencies can be situated in the upper complex plane $\omega''>0$.} and composed of two parts \cite{hanson2002operator}:
\begin{itemize}
 \item the discrete spectrum associated with \emph{proper eigenfunctions} known as
 trapped modes (also called bounded or guided modes) exponentially decreasing at infinity, 
 particularly the ``ideal'' surface plasmon modes when the structure contains materials with $\varepsilon<0$,
\item the continuous spectrum associated with \emph{improper eigenfunctions} composed of propagative or evanescent radiation modes.
\end{itemize}
In addition, another type of solution can be defined and is very useful to characterize the diffractive properties of unbounded structures: 
the so-called \emph{leaky modes}. These modes are an intrinsic feature of open waveguides. 
The associated eigenfrequencies are complex solutions of the dispersion relation of the problem but are \emph{not} 
eigenfrequencies of (\ref{eq:eigenpb}). A leaky mode represent the analytical continuation of the proper discrete mode 
below its cutoff frequency \cite{hanson2002operator}.\\
PMLs have proven to be a very convenient tool to 
compute leaky modes
in various configurations \cite{pmlCompel,Popovic2003,HEIN2004,Eliseev2005}. Indeed they mimic
efficiently the infinite space provided a suitable choice of their parameters. 
We may define a transformed operator with infinite PMLs, namely $\Mop_{\tens{\xi}^\mathrm{s}}$, with equivalent 
material properties defined by Eq.(\ref{equ_tranform}). The associated spectral problem is:
\begin{equation}
\Mop_{\tens{\xi}^\mathrm{s}}(v^\mathrm{s}_n):=-\div(\tens{\xi}^\mathrm{s}\,\grad v^\mathrm{s}_n)=\Lambda^\mathrm{s}_n \,\chi^\mathrm{s} \, v^\mathrm{s}_n.
\label{eq:eigenpb_transformed}
\end{equation}
Figure~\ref{PML_leaky} 
shows how the spectrum of the considered operator
is affected by applying a complex stretch in the non periodic case (See Appendix~\ref{appendix:loc_TCS} for more details). 
The introduction of infinite PMLs rotates the continuous spectrum 
in the complex plane (since the operator $\Mop_{\tens{\xi}^\mathrm{s}}$ involved in the problem is now a non self-adjoint extension of the 
original self-adjoint operator $\Mop_{\tens{\xi}}$).
The effect is not only to turn the continuous spectrum
into complex values but it also unveils the leaky modes is the region swept by the rotation 
of this essential spectrum \cite{thesebenjg}. It is important to note that leaky modes do \emph{not} 
depend on the choice of a particular complex stretching: adding the infinite PMLs is only a way to 
discover them. The angle of rotation of the continuous spectrum in $\mathbb{C}$ is the opposite of 
the argument $\phi$ of the constant complex stretching coefficient $s_\eta$. By increasing this parameter we discover more and more leaky modes with now exponential decay 
at infinity in the PML regions, and so the associated norms become finite.\\
Finally, the PMLs can safely be truncated at finite 
distance which results in an operator $\Mop_{\tens{\xi}^\mathrm{t}}$ having only discrete spectrum, which leads to the spectral problem:
\begin{equation}
\Mop_{\tens{\xi}^\mathrm{t}}(v^\mathrm{t}_n):=-\div(\tens{\xi}^\mathrm{t}\,\grad v^\mathrm{t}_n)=\Lambda^\mathrm{t}_n \,\chi^\mathrm{t} \, v^\mathrm{t}_n.
\label{eq:eigenpb_transformed_trunc}
\end{equation}
This formulation in the form of an equivalent transformed closed problem allows the numerical computation with the FEM
of approximate leaky, guided and radiation 
modes (also termed as PML modes or B\'erenger modes). This last set of modes is due to the discretization of the continuous
spectrum by finite PMLs \cite{Olyslager2004} with constant stretch and by the spatial discretization of the domain with a mesh
 in the framework of the FEM. The discretization of the continuous
spectrum is finer when either the thickness of the PMLs 
or the modulus $\sigma$ of the complex stretching coefficient $s_\eta$ increase. 
The boundary conditions and the FEM setup are analogous to that described in section \ref{subsection:eigpb}. 
Note that Neumann or Dirichlet boundary conditions applied in the outward boundaries of the PMLs result in a different set of approximate radiation modes.
 Obviously, leaky modes do \emph{not} depend on all those PML-related parameters.\\
The final algebraic system can be written in a matrix form as a generalized eigenvalue problem $A\,v=\Lambda B\,v$.
 Finding the eigenvalues closest to an arbitrary shift $\Lambda^0$ boils down to compute the largest 
 eigenvalues of matrix $C=(A-\Lambda^0 B)^{-1}B$. For this purpose, the eigenvalue solver uses 
 ARPACK FORTRAN libraries adapted to large scale and sparse matrices \cite{Lehoucq97arpackusers}. This code is based on a 
 variant of the Arnoldi algorithm called \textit{Implicitly Restarted Arnoldi Method} (IRAM).\\
 
In the sequel we will drop the exponent $\mathrm t$ for convenience, but one shall bear in mind that 
the effective problem we are dealing with is the complex stretched and bounded version (\ref{eq:eigenpb_transformed_trunc}) 
of the original problem (\ref{eq:eigenpb}) defined on a whole real Cartesian an unbounded space.

\section{Quasimodal expansion method}\label{sec:mm}

\subsection{Inner product and adjoint eigenproblem}
For Hermitian problems, eigenvectors form a complete set of $L^2(\Omega)$
and every solution of the problem with sources can be expanded on this basis. 
But in the general case, the problem may be non self adjoint, and we lack the nice properties of 
Hermitian systems.
Nevertheless, we describe here a procedure to obtain an expansion basis of the solution space.
 For this we use the classical inner product of two functions $f$ and $g$ of $L^2(\Omega)$:
\begin{equation}
  \bra  f\mb g\ket :=\int_{\Omega} f(\B r)\, \conj{g}(\B r) \; \ddroit\B r.
\end{equation} 
Unlike self-adjoint problems, $\bra \chi v_n \mb v_m\ket \neq\delta_{nm}$, in other 
words the eigenmodes $v_n$ are not orthogonal with respect to this standard definition. This is the reason why we 
consider an adjoint spectral problem with eigenvalues $\conj{\Lambda_n}=(\conj{\omega_n}/c)^2$
and eigenvectors $w_n$. The adjoint operator $\Mop^\dagger_{\tens{\xi}}$ is defined by
\begin{equation}
\bra \vphantom{ \Mop^\dagger_{\tens{\xi}}(w)}\Mop_{\tens{\xi}}(v) \mb w \ket =\bra  v \mb \Mop^\dagger_{\tens{\xi}}(w)\ket 
\end{equation}
with the \emph{same} boundary conditions as the direct spectral problem, 
and is such that $\Mop^\dagger_{\tens{\xi}}=\Mop_{\tens{\xi}^\star}$ (See Appendix \ref{appendix:prop_adj} for the proof), 
where $A^\star={\conj{A}}^\mathrm{T}$ is the conjugate transpose of  matrix $A$.
The associated adjoint problem that we shall solve is (cf. Appendix \ref{appendix:prop_adj}):
\begin{equation}
\Mop^\dagger_{\tens{\xi}}(w_n)=\Mop_{\tens{\xi}^\star}(w_n)=-\div(\tens{\xi}^\star \,\grad w_n)= \conj{\Lambda_n}\,\conj{\chi}\, w_n.
\label{eq:eigenpb_adjoint}
\end{equation}

We know from spectral theory that the eigenvectors $v_n$ are bi-orthogonal to their adjoint 
counterparts $w_n$ \cite{hanson2002operator}:
\begin{equation}
 \bra  \chi v_n \mb w_m\ket =\int_{\Omega}\chi(\B r)\, v_n(\B r)\, \conj{w_m}(\B r) \; \ddroit\B r=K_n\delta_{nm}.
\label{eq:biortho}
\end{equation} 
where the complex-valued normalization coefficient $K_n$ is defined as  
\begin{equation}
K_n:= \bra  \chi v_n \mb w_m\ket =\int_{\Omega}\chi(\B r)\, v_n(\B r)\, \conj{w_n}(\B r) \; \ddroit\B r.
\end{equation} 
\subsection{Quasimodal expansion of the diffracted field}

Relation (\ref{eq:biortho}) provides a complete bi-orthogonal set to expand every field solution
of Eq.~(\ref{equ:u2d}) propagating in the open waveguide as:
\begin{equation}
  u_2^d(\B r,\omega)=\sum_{n=1}^{+\infty}P_n(\omega)\, v_n(\B r) + \int_{\Gamma_c} P_\nu(\omega)\, v_\nu(\B r) \; \ddroit\nu,
\label{decompmod_integrale}
\end{equation}
where $\Gamma_c$ is the continuous spectrum (a curve, with possibly a denombrable set of branches in the complex plane). 
The discrete coefficients $P_n$ and the continuous density $P_\nu$ are given by similar expressions:
\begin{equation}
 P_j(\omega)=\frac{1}{K_j}\bra  \chi u_2^d  \mb  w_j\ket =\frac{J_j(\omega)}{{\omega}^2-\omega_j^2},\quad  j=\{n,\nu\}
 \label{equPn}
\end{equation}
with
\begin{equation}
J_j(\omega)=\frac{c^2}{K_j}\bra \source_1  \mb   w_j \ket =\frac{c^2}{K_j}\int_{\Omega_{g'}}\source_1(\B r,\omega)\, \conj{w_j}(\B r) \;\ddroit\B r,
\label{equJn}
\end{equation}
where the integration is \emph{only performed on the inhomogeneities} $\Omega_{g'}$ since the 
source term $\source_1$ is zero elsewhere. Note that the last integral has to be taken in the distributional meaning 
which leads to a surface term on $\partial\Omega_{g'}$ because of the spatial derivatives in $\source_1$.\\
We are thus able to know how a given mode is excited when changing the incident field.
This modal expansion can be approximated by a discrete sum since the spectrum of the final
operator we solve for
involves only discrete eigenfrequencies, and in practice only a finite number $M$ of modes is 
retained in the expansion, so that we can write:
\begin{equation}
 u_2^d(\B r,\omega)\simeq\sum_{m=1}^{M}P_m(\omega)\, v_m(\B r).
\end{equation}

This leads to a reduced modal representation of the field which is well adapted when studying
the resonant properties of the open structure, as illustrated in the sequel.\\

Equation (\ref{equPn}) shows clearly that the complex eigenfrequency $\omega_n$ is a simple pole of the coupling coefficient $P_n$ and thus leads to a singularity of the diffracted 
field. But in practice, the frequency of the incident plane wave is real, and the resonant behavior may happen in the vicinity of $\omega'_n=\re(\omega_n)$. 
Consequently, the value of $P_n$ is finite, and the linewidth of the resonance is given by $\omega''_n=\im(\omega_n)$. This is the main strength of the QMEM: 
it unambiguously reveals not only that a mode is excited but it indicates also the intensity of this excitation. According to Eq.~(\ref{decompmod_integrale}), 
one can see that the diffracted field for a given incident frequency is due to the concomitant contributions of an infinity of eigenmodes. 
However, for a given incident field, there is often a mode that plays a leading
role in the decomposition. In other words, its coupling coefficient is much larger in module than those associated with other modes, and so 
a resonance of the diffracted field may be attributed mainly to the excitation of this mode.

\subsection{Green function and Local Density Of States}

We have focused our attention on a plane wave source, but the method is 
also applicable for other type of excitation. Indeed, if we assume a point source 
located at $\B{r'}$, namely $\source_1(\B{r})=\delta(\B{r}-\B{r'})$, we have from Eq.~(\ref{equJn})
$$J_n=\frac{c^2}{K_n}\conj{w_n}(\B{r'}),$$
so we obtain immediately the Green function expansion in terms of quasimodes and adjoint quasimodes as : 

\begin{equation}
 g(\omega,\B{r},\B{r'})=\sum_{m}\frac{c^2}{K_m}\frac{v_m(\B{r})\,\conj{w_m}(\B{r'})}{\omega^2-\omega_m^2}.
 \label{equGF}
\end{equation}

 The Local Density Of States (LDOS) defined as 
 $$l(\omega,\B{r})=-\frac{2\,\omega}{\pi\,c^2}\,\im \left\lbrace g(\omega,\B{r},\B{r})\right\rbrace$$ can thus be expanded as : 
 
\begin{equation}
l(\omega,\B{r})=-\frac{2\,\omega}{\pi}\sum_{m}  \im\left\lbrace \frac{v_m(\B{r})\,\conj{w_m}(\B{r})}{K_m(\omega^2-\omega_m^2)}\right\rbrace.
 \label{equLDOS}
\end{equation}

The LDOS is thus related to \emph{local} values of eigenvectors and adjoint eigenvectors conjugates. 
Note that the QMEM is in this case highly computationally efficient, since it only requires to 
solve two spectral problems with the FEM to obtain the LDOS in a given region of space, without the need 
to compute numerically the integrals in Eq.~(\ref{equJn}). Once the eigenmodes of the system and their adjoints 
have been computed, the calculation of the LDOS at any point in the computational domain and at any frequency is trivial.
 This has to be compared with the resolution of a large number of direct FEM problem where 
 the source point position and the frequency vary.

\section{Numerical examples}\label{sec:exnum}

\subsection{Triangular rod in vacuum}\label{ssec:exnumtri}

\begin{figure}[htbp!]
\includegraphics[width=0.9\columnwidth]{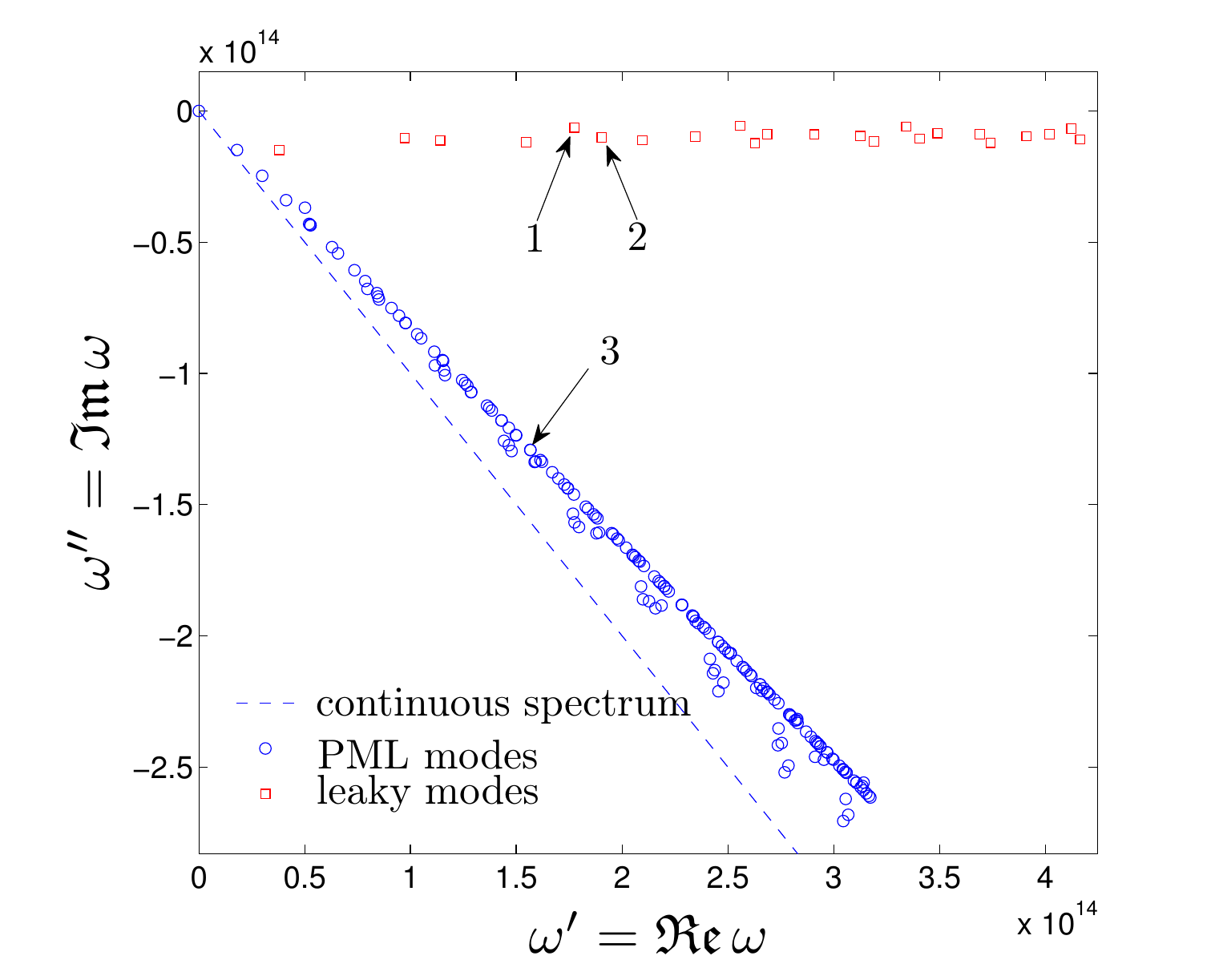}
\caption{Loci of the eigenfrequencies in the complex $\omega$-plane. Theoretical 
continuous spectrum (blue dashed line) is well approximated by discrete eigenvalues corresponding
to PML modes (blue circles). The leaky modes unveiled by shifting the continuous spectrum 
in the complex plane have frequencies represented by red squares.\label{plan_C_triangle}}
\end{figure}

\begin{figure*}[htbp!]
\includegraphics[width=0.7\linewidth]{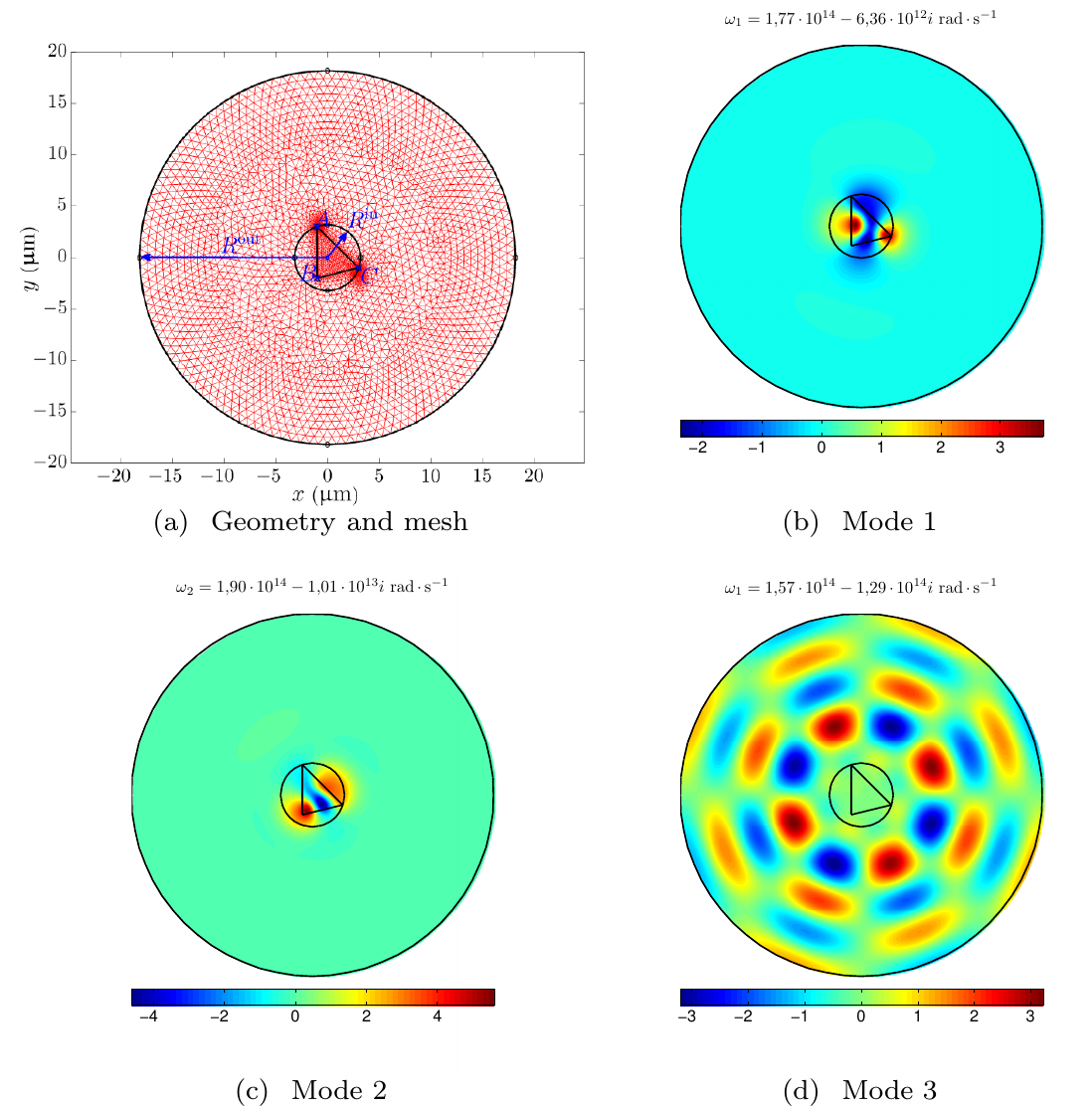}
\caption{Geometry and mesh of the structure (a) and field maps $\re(E_z)$ for the eigenmodes 1 (b), 2 (c) and 3 (d).}
\label{modes_triangle}
\end{figure*}

The first example is the case of a dielectric rod ($\varepsilon_{g'}=13-0.2\ic$ and $\mu_{g'}=1$) of infinite extension 
along the $z$-axis embedded in vacuum (See Fig.~\ref{modes_triangle}(a)). Its cross section is a triangle defined 
by the three apexes A $(-1;3)$, B $(-1;-2)$ and C $(3,-1)$. We chose the inner radius of the
PML to be $R^{\rm{in}}=1.01\cdot\rm{max}(OA,OB,OC)$, \textit{i.e.} to put the PML as 
close as possible to the diffractive object to avoid numerical pollution of the results as reported by previous studies \cite{KP09}. The depth
of the PML annulus is $R^{\rm{out}}-R^{\rm{in}}=\SI{15}{\micro\meter}$, and the absorption coefficient
is $s_r=1+\ic$ (cf. Eq. (\ref{ys})). We solve the eigenproblem in TE polarization, and the position of the 300 eigenfrequencies 
with lowest real parts in the complex plane is shown in Fig.~\ref{plan_C_triangle}. The original continuous
spectrum (for the problem without PML) is $\mathbb{R}^+$. It is rotated of an angle 
$\phi=-\mathrm{arg}( s_r)=-\pi/4$ from the real axis when using PML (blue dotted curve). 
The truncation of PML at a finite distance results in a discrete approximation of this continuous 
spectrum (blue circles).

The field of the associate quasi 
radiation modes is concentrated mainly in the PML region, as can be seen from the field map 
of mode 3 plotted in Fig.~\ref{modes_triangle}(d). Eigenvalues corresponding to leaky modes are
situated closest to the real axis (red squares), and the field profiles of the associated modes are 
confined in the region of physical interest $r<R^{\rm{in}}$ (See Figs.~\ref{modes_triangle}(b) and \ref{modes_triangle}(c) for leaky modes 1 and 2 respectively).\\

\begin{figure}[htbp!]
\includegraphics[width=0.9\columnwidth]{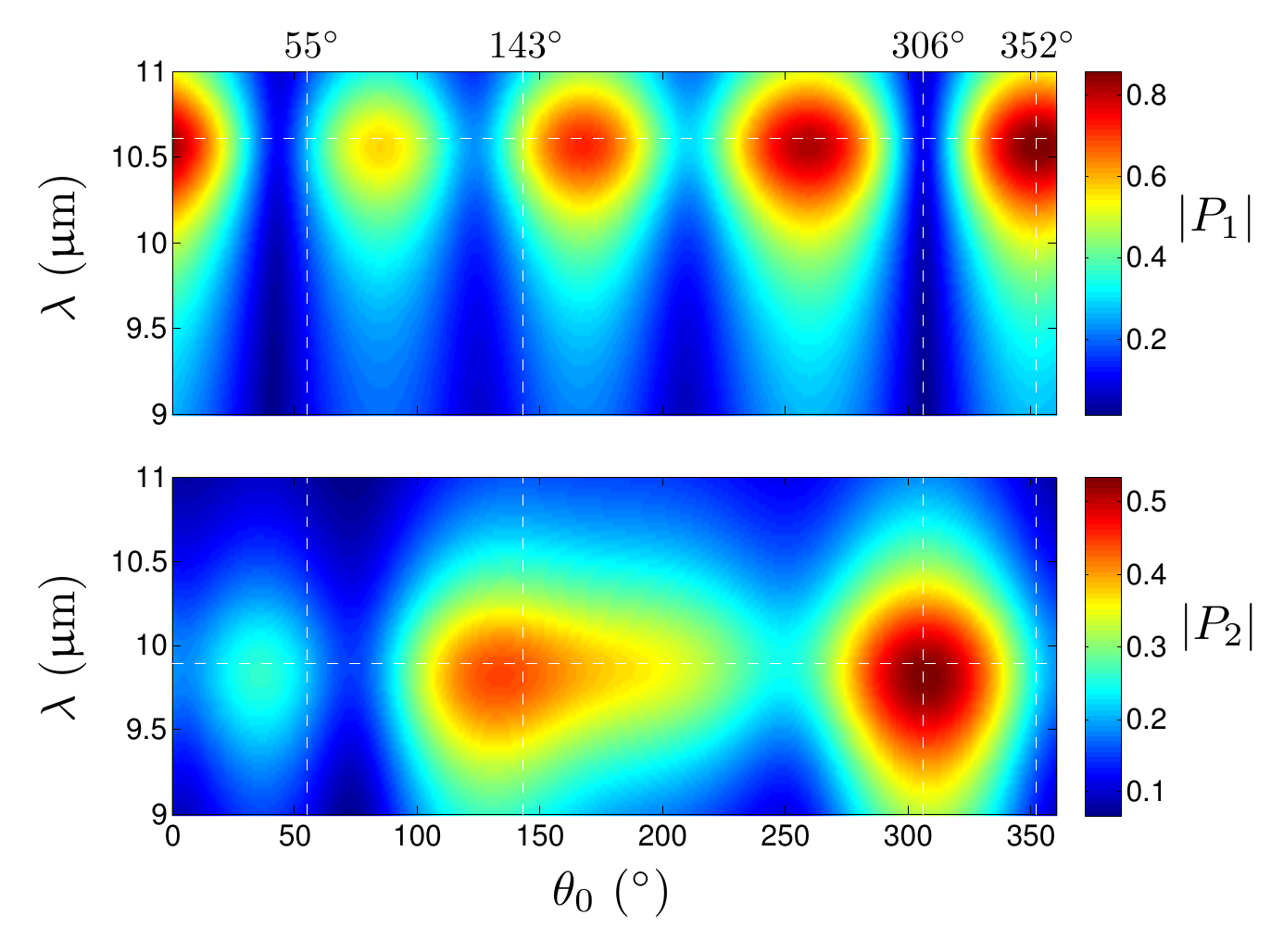}%
\caption{Coupling coefficients $P_n$ as a function of $\lambda$ and $\theta_0$ for 
the modes $1$ (top) and $2$ (bottom). Horizontal dashed lines indicate the resonant 
wavelength.\label{Pn_triangle}}
\end{figure}

We focus on two leaky modes labeled $1$ and $2$ for which associated eigenfrequencies are 
respectively $\omega_1= (\num{1.77e13} - \num{6.36e11}\,\ic)\,\si{\radian\per\second}$ (resonant wavelength
$\lambda_1=\SI{10.61}{\micro\meter}$) and $\omega_2= (\num{1.90e13} - \num{1.01e12}\,\ic)\,\si{\radian\per\second}$
($\lambda_2=\SI{9.89}{\micro\meter}$). 
In order to understand how these eigenmodes are excited, we compute the modal coefficients $P_n$ for varying incident 
wavelength $\lambda$ and angle $\theta_0$. The maps of the modulus of $P_n$ ($n=1,2$) for $\lambda$ between 9 
and $\SI{11}{\micro\meter}$ and $\theta_0$ between $0$ and $\SI{360}{\degree}$ are plotted in Fig.~\ref{Pn_triangle}. The coupling
coefficients $P_n$ behave as $1/(\omega^2-\omega_n^2)$, which yields a resonant behavior when $\omega$ is near
$\re(\omega_n)$ (cf. the horizontal dashed lines in Fig.~\ref{Pn_triangle}). We observe that the values of $\abs{P_n}$ strongly depend on
$\theta_0$, indicating that the considered mode will be more or less excited depending of the incidence. \\

\begin{figure}[htbp!]
\includegraphics[width=0.9\columnwidth]{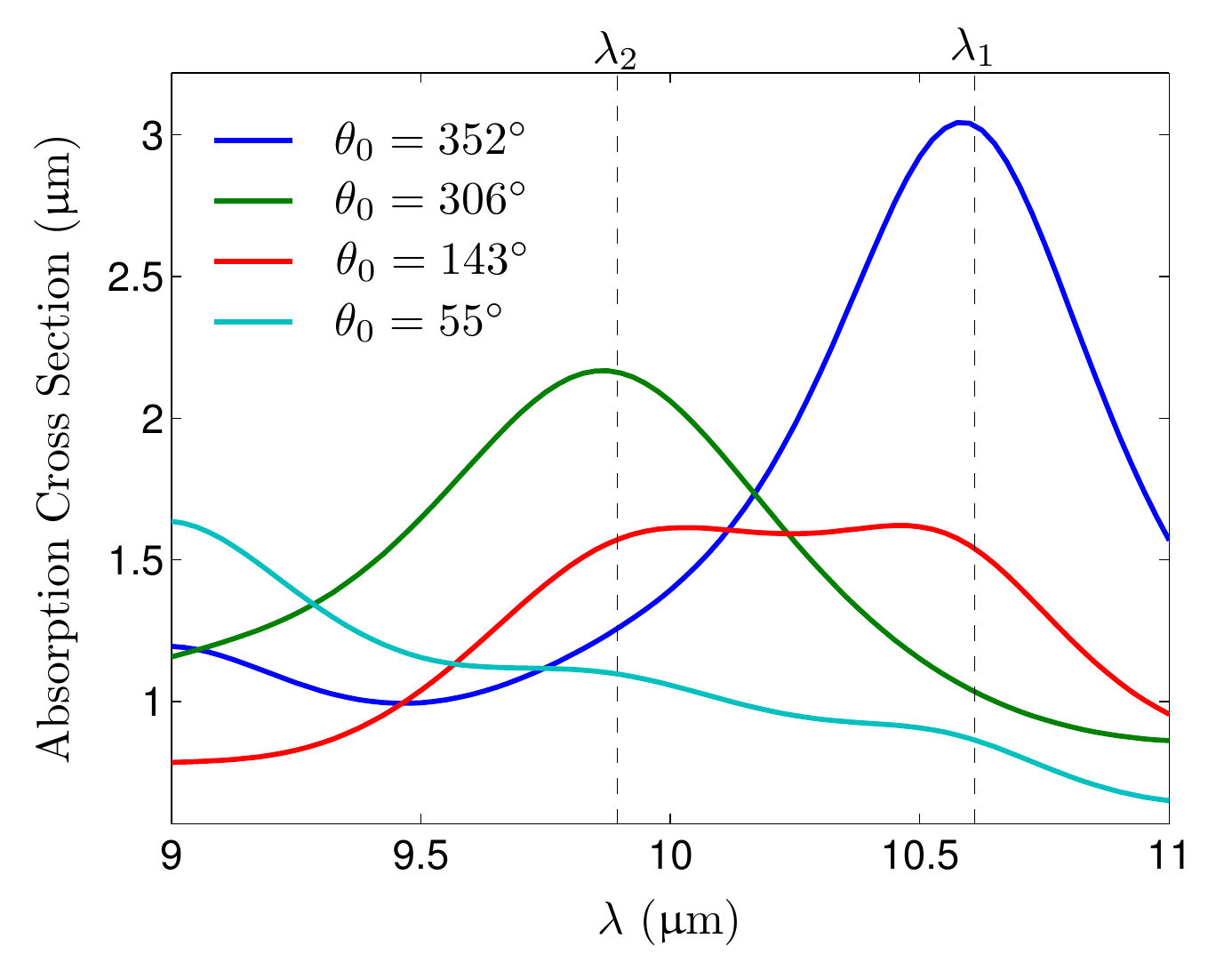}%
\caption{Absorption cross section as a function of $\lambda$ for different incident 
angles $\theta_0$.\label{ACS_triangle}}
\end{figure}

To check our previsions, we 
compute the absorption cross section by the method presented in part \ref{part:scatt_pb} at 
different incidences. In the first case where $\theta_0=352$\si{\degree}, the value of $\abs{P_1(\lambda_1)}$
is high whereas the value of $\abs{P_2(\lambda_2)}$ is much lower, which means that the mode 1 will be
principally excited. This is what can be seen on Fig~\ref{ACS_triangle} (blue curve) where the 
resonant peak of the absorption cross section curve occurs near $\lambda_1$, whereas no significant
resonant behavior is found near $\lambda_2$. Similar conclusions can be made 
for the second case with $\theta_0=306$\si{\degree}~by interchanging the roles of modes 
1 and 2. Note that the resonant peak in the second case is broader since $\im(\omega_2)>\im(\omega_1)$,
in other words the mode 2 leaks more than the mode 1. In addition, the value of the absorption cross 
section at the resonance is correlated to the value of the coupling coefficient $P_n$ for the corresponding
excited eigenmode: the peak value in the first case is 
greater because the mode is more excited comparing to the second case. Another 
interesting example is when the two modes have comparable weight in the modal 
expansion. This is the case for $\theta_0=143$\si{\degree}, so that both modes are excited.
In our case, the two resonant peaks in the absorption cross section curve merge into a 
single broad one (See the red curve on Fig~\ref{ACS_triangle}). Finally for $\theta_0=55$\si{\degree}, both 
modes show weak coupling coefficients, which results in a relatively flat behavior of the absorption
cross section (cyan curve on Fig~\ref{ACS_triangle}). In fact another mode dominates in this 
case with resonant wavelength slightly lower than $\SI{9}{\micro\meter}$.\\

\begin{figure}[htbp!]
\includegraphics[width=0.9\columnwidth]{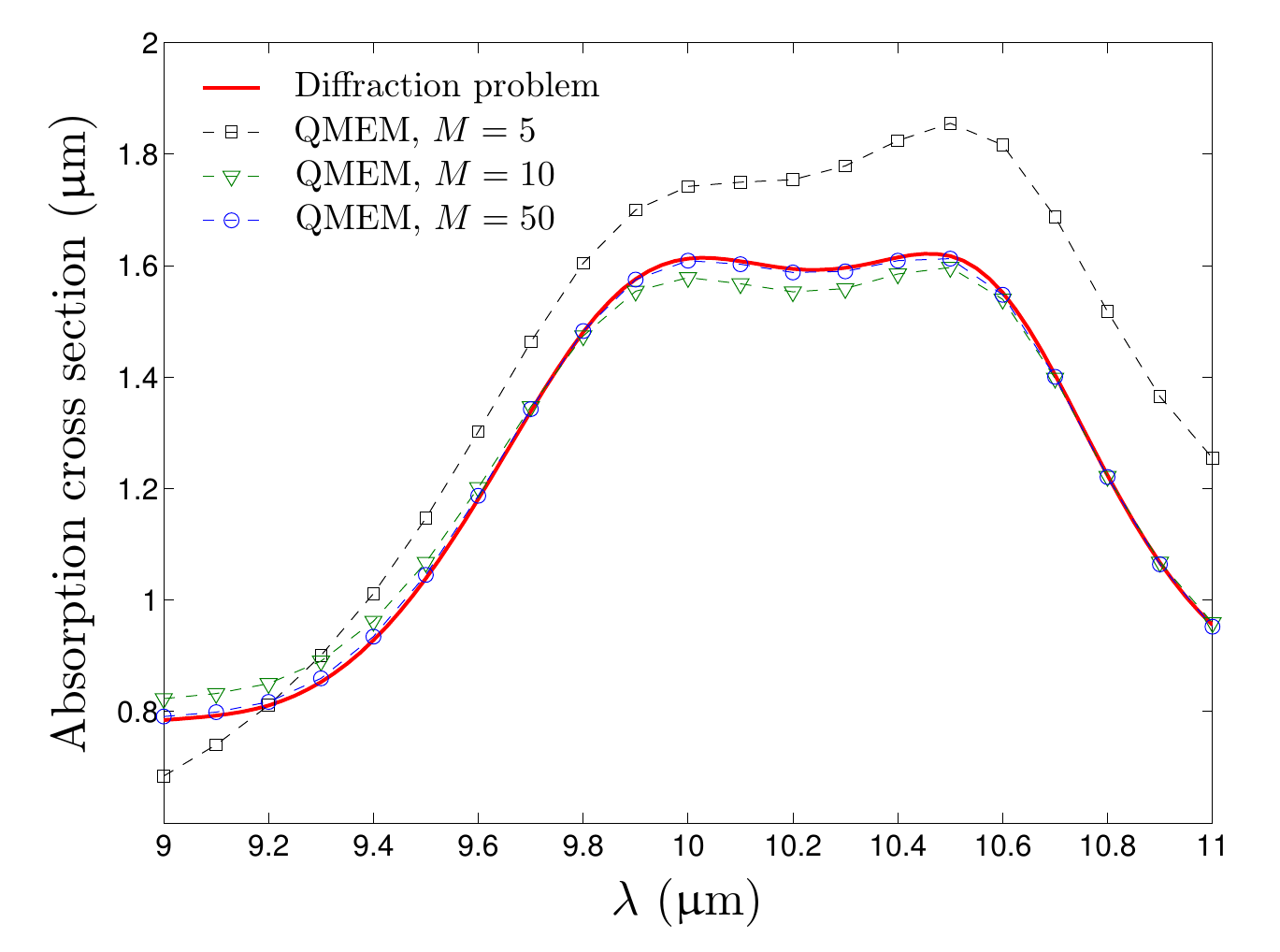}%
\caption{Absorption cross section curves computed with QMEM as a function of $\lambda$ for $\theta_0=143$\si{\degree}. 
The thick red curve corresponds to the reference values computed by solving the diffraction problem.\label{rec_triangle}}
\end{figure}

\begin{figure}[htbp!]
\includegraphics[width=0.9\columnwidth]{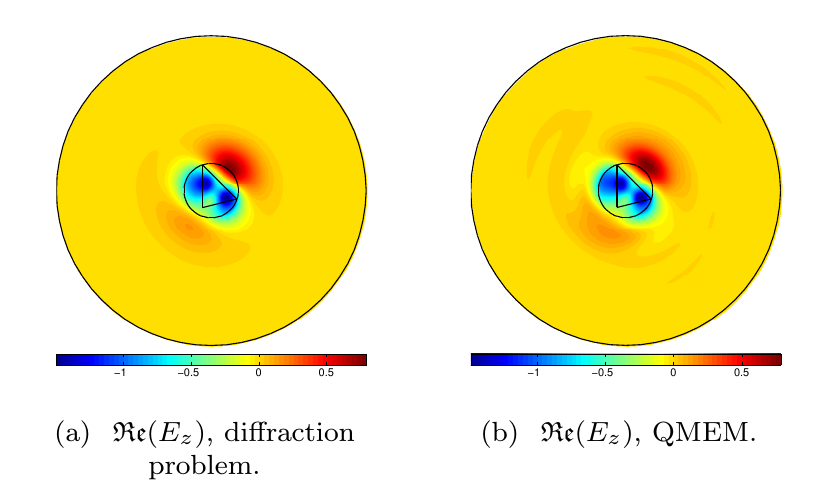}%
\caption{Electric field at $\lambda=\SI{10.2}{\micro\meter}$ and $\theta_0=\SI{143}{\degree}$ calculated 
by solving the diffraction problem (a) and by the QMEM (b) with 50 modes.}
\label{champs_reconstruits_triangle}
\end{figure}

Another powerful feature of our approach is that we are able to reconstruct the 
field with a few eigenmodes.
 From this reduced modal expansion we calculate the absorption cross section for 
 $\theta_0=143$\si{\degree}. The $M$ modes used are those with highest mean value of 
 the modal coefficient on the whole wavelength range. Results are reported on Fig~\ref{rec_triangle} and
 compared with the reference values obtained by solving Eq.~(\ref{equ:u2d}). For $M=5$, we have 
 already captured evolution of the absorption cross section with frequency . The agreement is better
 for $M=10$ except for weak wavelengths. Retaining $M=50$ modes in the modal expansion
 results in an accurate approximation of the absorption cross section. We plot in
 Fig.~\ref{champs_reconstruits_triangle} the field maps obtained by solving the 
 diffraction problem and the modal method approximation with 50 modes, 
 at $\lambda=\SI{10.2}{\micro\meter}$ and $\theta_0=143$\si{\degree}. As can be seen, 
 the two methods are in good agreement, with only local discrepancies occurring at 
 the interface air/PML and within the PML. Note that this reduced order model is 
computationally efficient when a large range of incident parameters is investigated. 
Indeed, there is only one FEM problem solved for (because in that case the adjoint modes $w_n$ 
are simply the conjugate of the eigenmodes $v_n$, See Appendix \ref{appendix:prop_adj} for the proof), 
the rest of the calculation is only numerical integration of smooth functions and algebraic operations.\\

\begin{figure}[htbp!]
\includegraphics[width=0.9\columnwidth]{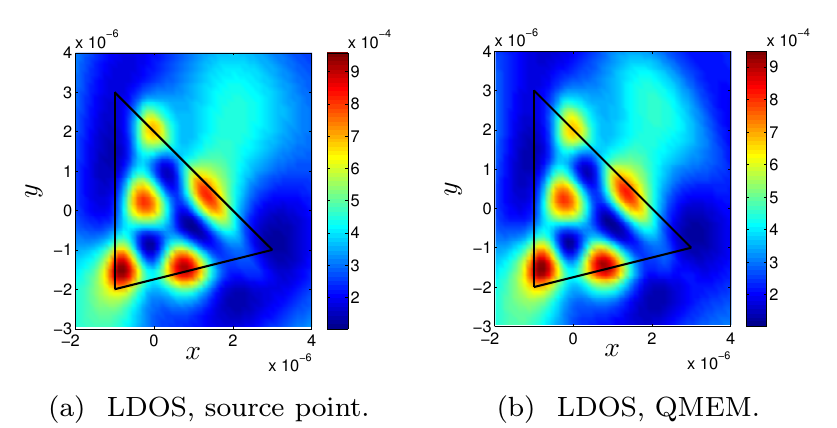}%
\caption{Local Density Of States at $\lambda=\SI{10.2}{\micro\meter}$ calculated 
by solving the diffraction problem (a) and by the QMEM (b) with $M=$500 modes.}
\label{LDOS_triangle}
\end{figure}

Finally, we computed a map of the LDOS at $\lambda=\SI{10.2}{\micro\meter}$ on a regular grid with $50\times 50$ points 
into the spatial window $[-2,4]\,\si{\micro\meter}\times[-3,4]\,\si{\micro\meter}$ around the dielectric rod.
 The results of the QMEM using Eq. (\ref{equLDOS}) with $m=500$ (See. Fig~\ref{LDOS_triangle}(b)), that involves the resolution 
 of a single FEM spectral problem, is in excellent agreement with the
 results obtained by 2500 direct FEM problems where the position of the source varies on the nodes of the $50\times 50$ grid (See. Fig.~\ref{LDOS_triangle}(a)). 
 In that example, the spectral problem consisting of \num{11753} degrees of freedom was solved in \SI{17}{min} on a laptop with 
 two \SI{2.8}{GHz} processors and \SI{8}{Go} of RAM. On the one hand, the computation of the modes is the limiting step but afterward the 
 LDOS are calculated in approximately one second. On the other hand, the computation of the LDOS on the $50\times 50$ grid with the direct problem takes more than one hour.
 Moreover the LDOS can be calculated at other wavelengths without any need of additional time consuming FEM simulations: 
 for 50 wavelengths the direct problem would take more than two days whereas it takes less than one minute with the QMEM (once the modes have been computed). 
 This example shows clearly the numerical efficiency of the QMEM compared to direct simulations.

\subsection{Lamellar diffraction grating}

\begin{figure}[htbp!]
\includegraphics[width=0.97\linewidth]{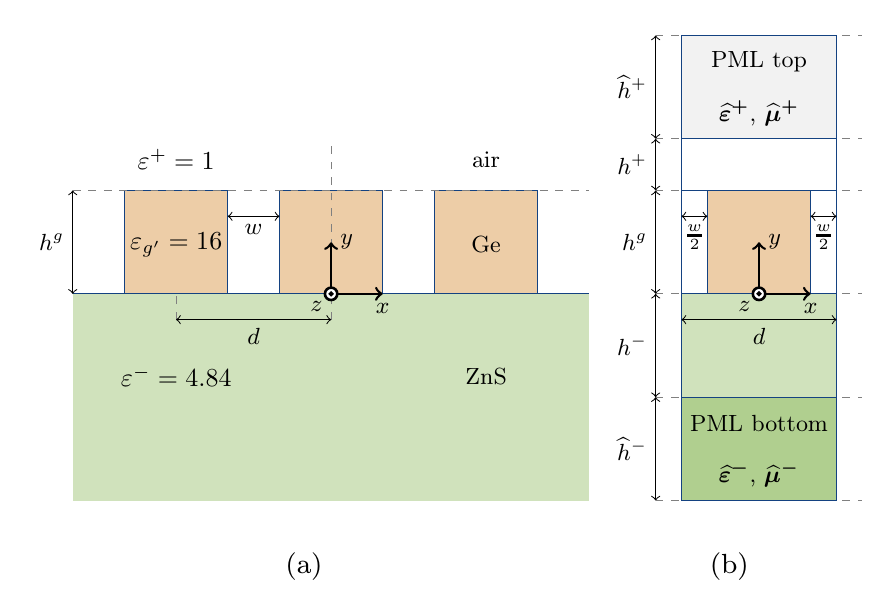}
\caption{Setup of the problem for the lamellar grating. 
(a): sketch of the studied diffraction grating. Parameters are 
 $w=\SI{0.1}{\micro\meter}$, $h^g=\SI{3}{\micro\meter}$, $\varepsilon_{g'}=16$, $\varepsilon^-=4.84$, $\varepsilon^+=1$, $d=\SI{3}{\micro\meter}$, 
 all materials are non magnetic ($\mu=1$.) 
 (b): computational cell for the FEM calculations. Top and bottom PML have stretching coefficient
 $\zeta^+=\zeta^-=\e^{\ic\frac{\pi}{4}}$ and their thicknesses are $\widehat{h}^\pm=\lambda^{\rm ref}/\sqrt{\varepsilon^\pm}$. 
 The thicknesses of the substrate and superstrate are $h^\pm=\lambda^{\rm ref}/10$, with $\lambda^{\rm ref}=\SI{14}{\micro\meter}$. 
 We apply quasiperiodicity conditions on the lateral boundaries with $\alpha=\SI{0}{\radian\per\meter}$ and Neumann homogeneous boundary conditions 
 on the outward boundaries of the PML. Maximum mesh element size is set to be $\lambda^{\rm mesh}/(20\sqrt{\abs{\re(\varepsilon)}})$, 
where $\lambda^{\rm mesh}=\SI{11}{\micro\meter}$.\label{fig:schema_resfentes_MM}}
\end{figure}

We focus in this section on the periodic case. Let us consider a mono-periodic diffraction grating (See Fig.~\ref{fig:schema_resfentes_MM}) constituted
 of slits of width $w$ engraved in a germanium layer of permittivity $\varepsilon_{g'}=16$ and of thickness $h^g=\SI{3}{\micro\meter}$. 
 The grating is deposited on a ZnS substrate of permittivity $\varepsilon^-=4.84$ and the superstrate is air ($\varepsilon^+=1$). 
The computational cell is limited to a strip of width $d$ with quasiperiodicity conditions on the lateral boundaries of coefficient $\alpha$.
The substrate and superstrate are truncated by PML and their thicknesses are $h^\pm=\lambda^{\rm ref}/10$, with $\lambda^{\rm ref}=\SI{14}{\micro\meter}$. 
Top and bottom are PML terminated by Neumann homogeneous boundary conditions and have stretching coefficient are $\zeta^+=\zeta^-=\e^{\ic\frac{\pi}{4}}$. \\

\begin{figure}[htbp!]
\includegraphics[width=0.97\linewidth]{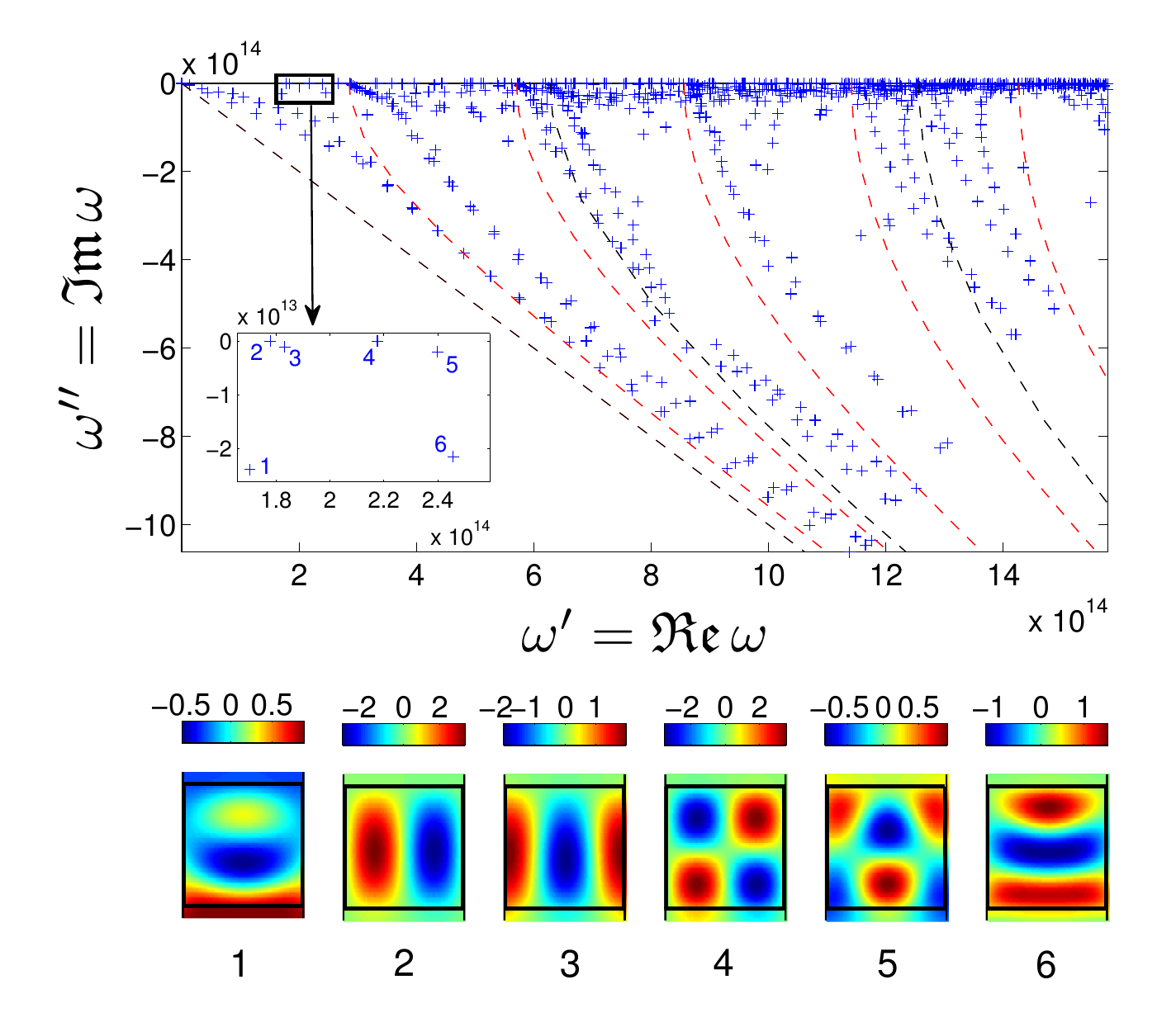}
\caption{Spectrum of the problem and leaky modes for $\alpha=\SI{0}{\radian\per\meter}$. Top: eigenfrequencies in the complex plane (blue crosses) 
and theoretical curves of the continuous spectrum (dashed red and black curves for the substrate and the superstrate respectively). 
The inset shows the position of the eigenvalues corresponding to the six leaky modes studied. 
 Bottom: real part of $H_z$ for these six leaky modes.}
\label{planC_grating}
\end{figure}

\begin{figure*}[htbp!]
\includegraphics[width=0.97\linewidth]{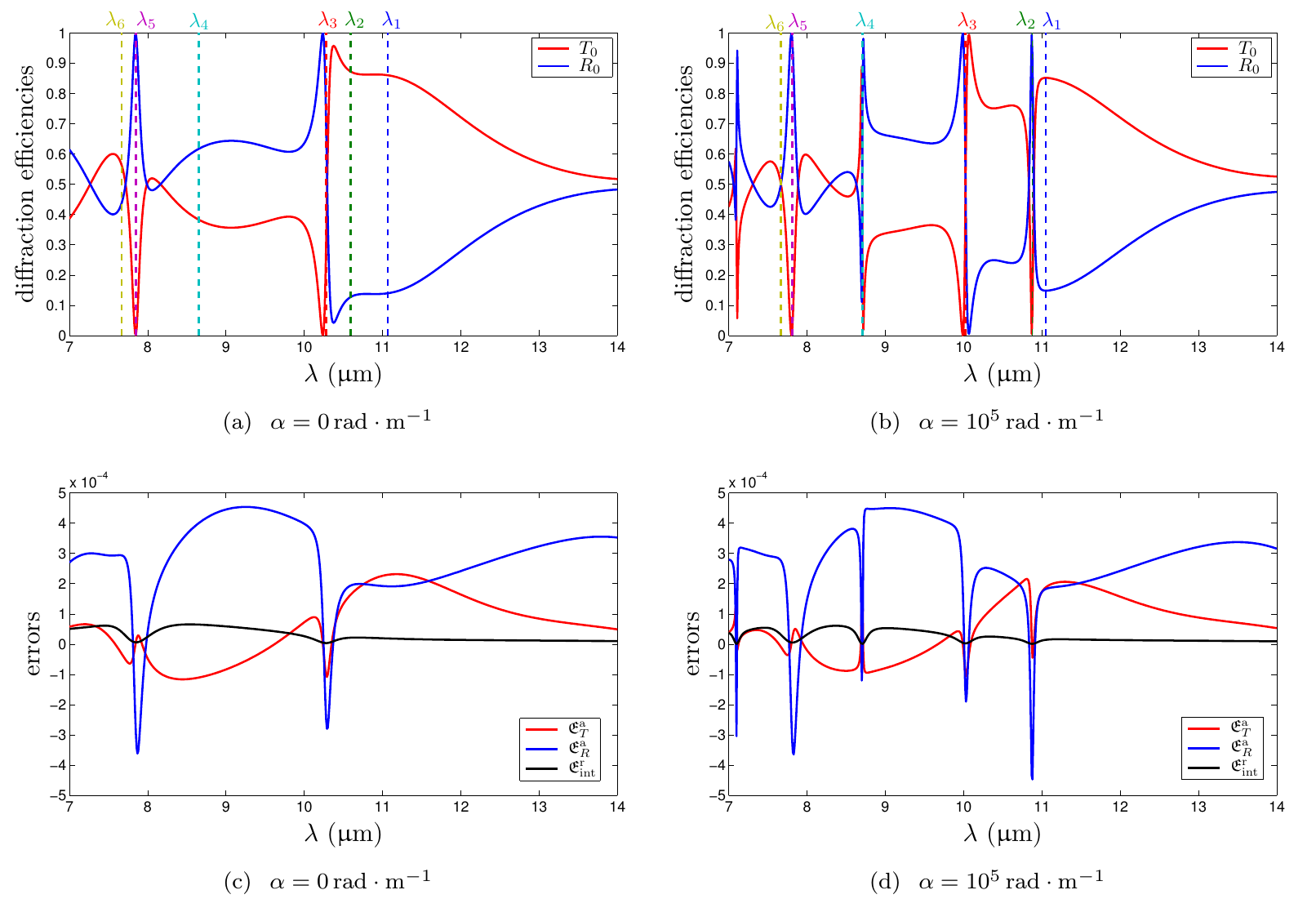}
\caption{Comparison between direct problem and QMEM. 
(a) and (b): reflection and transmission coefficients in the zeroth order $R_0$ and $T_0$. 
(c) and (d): relative integrated error $\mathfrak{E}^{\mathrm r}_{\mathrm {int}}$, absolute errors on transmission $\mathfrak{E}^{\mathrm a}_{T}$ 
and reflection $\mathfrak{E}^{\mathrm a}_{R}$.}
\label{b}
\end{figure*}

We computed the first 801 eigenfrequencies (with lowest real parts) of this grating for $\alpha=0$ and $\SI{e5}{\radian\per\meter}$, as well as their associated 
adjoints. The position of the eigenfrequencies in the complex plane as well as the theoretical curves of the continuous spectrum for $\alpha=\SI{0}{\radian\per\meter}$
are plotted on Fig.~\ref{planC_grating}. 
The deviation of the approximate radiation modes eigenfrequencies are due to the large grating-PML distance required to obtain an accurate result on the diffraction efficiencies, 
as we will see in the sequel. We focus on six leaky modes the resonant wavelength of which are in the far infrared spectral region $8-14\,\si{\micro\meter}$, corresponding to a transparency window of 
the atmosphere (See the inset in Fig.~\ref{planC_grating}). The field maps of those modes for $\alpha=\SI{0}{\radian\per\meter}$ are plotted on Fig.~\ref{planC_grating}. 
The corresponding resonant wavelength $\lambda_n=2\pi c/\omega'_n$ and quality factors $Q_n=\omega'_n/(2\omega''_n)$ are reported in Table \ref{table_modes}. \\
The modes labeled $1$ and $6$ have in both cases weak $Q$ factors, which means that the associated resonance is broad. 
This is confirmed by the observation of the diffraction efficiencies (Figs.~\ref{b}(a) and \ref{b}(b)), where a wide resonant peak is found around $\lambda_1$ and $\lambda_6$. 
For both values of $\alpha$, the resonant parameters of these low-$Q$ modes are almost unchanged. \\

\begin{table}[htbp!]
    \centering
\begin{tabularx}{\columnwidth}{>{\centering\arraybackslash}X>{\centering\arraybackslash}X>{\centering\arraybackslash}X>{\centering\arraybackslash}X>{\centering\arraybackslash}X}
 & \multicolumn{2}{c}{$\alpha=\SI{0}{\radian\per\meter}$} & \multicolumn{2}{c}{$\alpha=\SI{e5}{\radian\per\meter}$} \\ \hline\hline
{$n$} & $\lambda_n$ (\si{\micro\meter}) & $Q_n$ & $\lambda_n$ (\si{\micro\meter}) & $Q_n$\\\hline
{1} & \num{11.06} & \num{3.56} & \num{11.05} & \num{3.55}\\
{2} & \num{10.59} & \num{7.01e9} & \num{10.88} & \num{2.20e2}\\
{3} & \num{10.28} & \num{8.51e1} & \num{10.02} & \num{1.35e2}\\
{4} & \num{8.65} & \num{3.25e10} & \num{8.71} & \num{4.99e2}\\
{5} & \num{7.85} & \num{5.93e1} & \num{7.81} & \num{6.55e1}\\
{6} & \num{7.67} & \num{5.69} & \num{7.66} & \num{5.71}\\
\hline\hline
\end{tabularx} 
\caption{Resonant wavelengths $\lambda_n$ and quality factors $Q_n$ of the modes for $\alpha=0$ and $\SI{e5}{\radian\per\meter}$.\label{table_modes}}
\end{table}

\begin{figure*}[htbp!]
\includegraphics[width=0.97\linewidth]{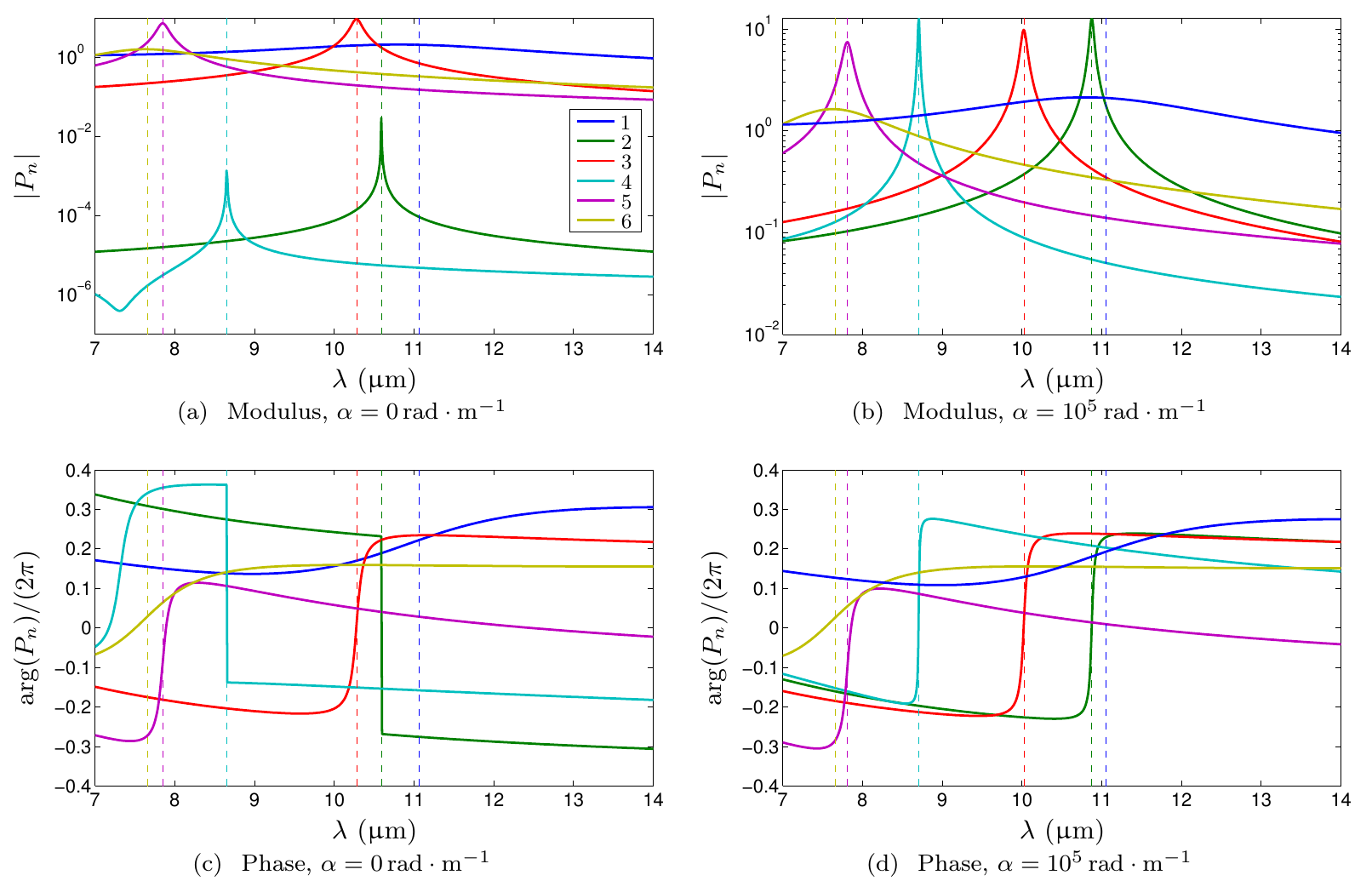}
\caption{Coupling coefficients $P_n$ as a function of the incident wavelength $\lambda$. 
The dashed vertical lines correspond to the position of the resonant wavelength $\lambda_n$ associated with each leaky mode.}
\label{Pna}
\end{figure*}

The coupling coefficients $P_n$ for the six leaky modes as a function of $\lambda$ were computed and are reported on Fig.~\ref{Pna}. One clearly sees a 
resonant peak of the modulus of $P_n$ (See Figs.~\ref{Pna}(a) and \ref{Pna}(b)) and a phase jump (See Figs.~\ref{Pna}(c) and \ref{Pna}(d)) around the resonant wavelength $\lambda_n$. As expected, the variations are all the more curt that 
the imaginary part of the eigenvalue is weak. These curves also show the relative contribution of the eigenmodes to the overall diffraction process. 
The two modes labeled $3$ and $5$ with high quality factors provoke sharp resonances in the transmission and reflection spectra (See Figs.~\ref{b}(a) and \ref{b}(2)). 
The high value of the modulus of their coupling coefficient $P_n$ clearly betrays their role in these resonances (See red and magenta curves on Figs.~\ref{Pna}(a) and \ref{Pna}(b)). 
On the contrary, modes $2$ and $4$, which have a huge $Q$ factor for $\alpha=\SI{0}{\radian\per\meter}$ (which means they are ``quasi normal'' modes) are very weakly excited in comparison to others modes 
on the whole spectral band excepted at the corresponding resonant wavelength (See cyan and green curves on Fig.~\ref{Pna}(a) where the modulus of the coupling coefficients is very weak). 
These findings explain why we do not observe significant resonances on the diffraction efficiencies around $\lambda_2$ and $\lambda_4$ (See Fig.~\ref{b}(a)): the associated leaky modes are 
not sufficiently excited. Actually, since these modes have extremely low leakage, they shall produce a very narrow resonance. We have computed the diffraction efficiencies 
around $\lambda_2$ and $\lambda_4$ with a finer wavelength step and encountered effectively extremely sharp resonances but with very weak
variations of the reflection and transmission coefficients (of the order of $\num{e-6}$).
For $\alpha=\SI{e5}{\radian\per\meter}$, the resonant wavelength of these two modes slightly increases comparing to the case $\alpha=\SI{0}{\radian\per\meter}$, while their $Q$ factor 
dramatically collapse (cf. Table \ref{table_modes}). The coupling coefficients are in this case of the same order of magnitude than the others modes (See cyan and green curves on Fig.~\ref{Pna}(b)), 
implying sharp scattering resonances in the reflection and transmission spectra (See Fig.~\ref{b}(b)) around $\lambda_2$ and $\lambda_4$. One can observe another sharp resonance at a 
wavelength slightly greater than $\SI{7}{\micro\meter}$, which is not studied here.\\

The particular example presented here illustrates 
the potential complexity of the diffractive process. Indeed, there is for example two close resonances around $\SI{7.8}{\micro\meter}$ that give raise to an hybrid resonance 
of the diffraction efficiencies due principally to a mixture of mode $6$ (with low $Q$ factor, yielding a broad resonance) and mode $5$ (with high $Q$ factor, sharp resonance). 
The computation of the complex eigenvalues indicates the presence of these modes and their associated resonant wavelength and linewidth,
but the QMEM allows us to go further by tracking the relative weight of these modes in the scattering process.\\

In order to assess the precision of our method, we have computed the absolute errors on the efficiencies calculated by 
solving the diffraction problem (DP) and by the QMEM: 

\begin{equation*}
 \mathfrak{E}^{\mathrm a}_{T}=T_0^{\rm DP}-T_0^{\rm QMEM}\quad\text{for transmission,}
\end{equation*}
and
\begin{equation*}
\mathfrak{E}^{\mathrm a}_{R}=R_0^{\rm DP}-R_0^{\rm QMEM}\quad\text{for reflection.}
\end{equation*}
We also calculated the integrated relative error on the computational cell $\Omega$ defined as:
\begin{eqnarray*}
 \mathfrak{E}^{\mathrm r}_{\mathrm {int}} &&=
\frac{\bra  u_2^{d,\rm DP}-u_2^{d,\rm QMEM}   \mb  u_2^{d,\rm DP}-u_2^{d,\rm QMEM}\ket }{\bra  u_2^{d,\rm DP} \mb  u_2^{d,\rm DP}\ket }\\
  &&=\frac{ \int_\Omega \left\lvert u_2^{d,\rm DP}(\B r)-u_2^{d,\rm QMEM}(\B r)\right\rvert^2 \ddroit\B r}{ \int_\Omega \left\lvert u_2^{d,\rm DP}(\B r)\right\rvert^2\ddroit\B r}.
\end{eqnarray*}

These errors are plotted as a function of $\lambda$ on Figs.~\ref{b}(c) and \ref{b}(d). One can see that the integrated relative error remains 
inferior to $\num{e-5}$, and that the absolute errors on the diffraction efficiencies is smaller in absolute value than $\num{5e-4}$, which shows the accuracy of 
the QMEM. The main drawback is that we have to take into account a sufficiently large number of modes (here $801$) to reconstruct correctly the field and hence the Fresnel coefficients. 
In comparison with the example studied in section \ref{ssec:exnumtri} where only $50$ modes reproduces the absorption cross section well, 
we must reconstruct the field very well in the substrate and superstrate to 
obtain a satisfying accuracy on the transmission and reflection by taking into account a large number of approximated radiation modes (associated with the continuous spectrum). 
On the contrary, the absorption being located into the diffractive object, a smaller number of leaky mode is sufficient to obtain a good approximation of the field inside the scatterer.\\

\section{Conclusion}

The quasimodal expansion method (QMEM) has been implemented and validated in planar and possibly periodic open electromagnetic systems 
with arbitrary geometries. The determination of eigenmodes and eigenvalues of those structures, based on the treatment of an 
equivalent closed problem with finite PML with the FEM, has been presented. Once the spectrum of Maxwell's operator have been computed, 
the solution of the problem with arbitrary sources can be expressed as a linear combination of eigenstates and the
expansion coefficients can be calculated with the help of adjoint eigenvectors. The method developed has been illustrated on 
numerical examples, showing both its capacity to perform a precise modal analysis and its accuracy. The first example of a triangular rod 
provides the conditions of excitation of a given mode by a plane wave by studying the coupling coefficients as a function of angle and wavelength. 
A reduced order model with a few modes is proven to well approximate the absorption cross section. The computation of the LDOS on a 2D spatial grid around the nanoparticle at 
an arbitrary wavelength is straightforward and computationally very efficient once the eigenmodes and eigenvectors have been calculated. The second numerical example of a lamellar 
diffraction grating illustrates the ability of the method to compute the eigenmodes of periodic media. The richness of the transmission and reflection spectra 
with coupled resonances is fully explained by the study of modal expansion coefficients. The precision of the method is demonstrated in comparison with a 
diffraction problem solved by the FEM. The extension of the QMEM to three dimensional structures, including bi-periodic grating, will be reported in future works.

\appendix
\section{Expression of the source term}
\label{appendix:source_term}
The source term of the equivalent radiation problem (\ref{equ:u2d}) is defined as:
\begin{eqnarray*}
\source_1&:=&-\Lop_{\tens{\xi},\chi}(u_1)=\Lop_{\tens{\xi},\chi}
(u_1)+\underbrace{\Lop_{\tens{\xi_1},\chi_1}(u_1)}_{=0}\\
&\;=&\Lop_{\tens{\xi_1}-\tens{\xi},\chi_1-\chi}(u_1).
\end{eqnarray*}
Since on the one hand $\tens{\xi}$ and $\tens{\xi_1}$ , and on the other hand $\chi$ and $\chi_1$
 are equal everywhere but into the inhomogeneity, one can see that the support of the sources is bounded by 
 this diffractive element. Let's now detail the expression of this source term. Classical transfer
 matrix calculus used in thin film optics (See for example Ref. \onlinecite{mcleod})
is employed to obtain closed form for $u_1$:
\begin{eqnarray}
\label{decomp_u1}
u_1(x,y)=&&u_0(x,y)+\mathrm{exp}(i \alpha x)\\
&&\times\begin{cases}
 r\;\mathrm{exp}(-i\beta^+y)\; &\mbox{for } y>0,\\
 v_n^c +v_n^p \; &\mbox{for } y_n<y<y_{n-1},\\
 t\;\mathrm{exp}(i\beta^+y)\; &\mbox{for } y<y_N,
\end{cases}\nonumber
\end{eqnarray}
for $1<n<N$, where
 \begin{equation}
\begin{array}{lcl}
v_n^p&=&u_n^p \;\mathrm{exp}(-i\beta_n(y-y_n)), \\
v_n^c&=& u_n^c \;\mathrm{exp}(i\beta_n(y-y_n)),
\end{array}
\label{decomp_u1_2}
 \end{equation}
with $\beta_n^2= k_n^2-\alpha^2$. This transfer matrix formalism provides the complex coefficient
 $u_n^p$ and $u_n^c$ together with the complex transmission and reflection coefficient $t$ and $r$ of the multilayer 
 stack. Exponents $p$ and $c$ indicate the propagative or counter-propagative
 nature of the plane waves $v_n^p$ and $v_n^c$. Knowing the expression of $u_1$ in the groove 
 region (with index $g$) and the linearity of the operator $ \Lop_{\tens{\xi},\chi}$, the source 
 term can be split into two contributions:

\begin{equation}
 \source_1=\source_1^p+\source_1^c,
\end{equation}
where
\begin{equation}
 \source_1^p=\Lop_{\tens{\xi_1}-\tens{\xi},\chi_1-\chi}(v_g^p)
\end{equation}
and
\begin{equation}
 \source_1^c=\Lop_{\tens{\xi_1}-\tens{\xi},\chi_1-\chi}(v_g^c).
\end{equation}

Finally we can obtain these terms under a more explicit form:

\begin{eqnarray}
 \source_1^p=&& u_g^p\left\lbrace
i\;\div
\left[
\left(
\tens{\xi}^{g} - \tens{\xi}^{g'}
\right)\, {\B k}^{g,p}\mathrm{exp}(i{\B k}^{g,p}\cdotp{\B r})
\right] \right. \nonumber \\
&&\left. +k_0^2\left(\chi^{g} - \chi^{g'}\right)\mathrm{exp}(i{\B k}^{g,p}\cdotp{\B r})
\right\rbrace
\end{eqnarray}
and
\begin{eqnarray}
 \source_1^c=&&u_g^c\left\lbrace
i\;\div
\left[
\left(
\tens{\xi}^{g} - \tens{\xi}^{g'}
\right)\, {\B k}^{g,c}\mathrm{exp}(i{\B k}^{g,c}\cdotp{\B r})
\right] \right. \nonumber \\
&&\left.+k_0^2\left(\chi^{g} - \chi^{g'}\right)\mathrm{exp}(i{\B k}^{g,c}\cdotp{\B r})
\right\rbrace
\end{eqnarray}
where ${\B k}^{g,p}$ (resp. ${\B k}^{g,c}$) is the wavevector associated
with the propagative (resp.~counter-propagative) wave in layer
$g$ as defined by equations (\ref{decomp_u1}) and (\ref{decomp_u1_2}).\\

\section{Location of the transformed continuous spectrum}
\label{appendix:loc_TCS}
We derive here the location of the continuous spectrum when adding infinite PMLs 
with constant coordinate stretching. 
Let us first consider a closed problem of a Fabry-P\'erot cavity of length $h$ with perfect conducting walls 
embedded in a homogeneous, lossless and isotropic medium of permittivity $\varepsilon$ and permeability $\mu$, the 1D-eigenproblem of which is :
\begin{align*}
 &  \mathcal{M}(v_n):=-\frac{\mathrm{d}^2v_n}{\mathrm{d}y^2}=\frac{\omega_n^2}{c^2}\varepsilon\mu v_n,\hspace{10pt} \forall y\in [0,h] \\
&v_n(0)=v_n(h)=0.
\end{align*}
The eigenvalues $\omega_n=n\,{\pi\,c}/(\sqrt{\varepsilon\,\mu}\,h)$, $\forall n\in\mathbb{N}^\star$, 
are real an positive and form discrete set as the problem is closed and self adjoint. Now if the problem is open ($h=+\infty$), one can 
see that the discrete set of eigenvalues $\omega_n$ tends to a continuous spectrum which proves to be $\mathbb{R}^+$. 
This result can be generalized to a class of problems known as singular Sturm-Liouville problems \cite{hanson2002operator}. \\

\subsection{The non periodic case with cylindrical PMLs}
In cylindrical coordinates $(\rho,\psi)$, we seek a separation of variables solution $v(\rho,\psi)=R(\rho)\Psi(\psi)$. 
The Helmholtz spectral equation for the variable $\rho$ reads the so-called radial Bessel equation: 
\begin{equation}
- \frac{1}{\rho} \frac{\ddroit }{\ddroit \rho}\left( {\rho} \frac{\ddroit R(\rho)}{\ddroit \rho}\right)
+\left(\frac{m^2}{\rho^2}-\varepsilon\mu k^2\right) R(\rho)  =0,
\label{eq_Bessel_rad}
\end{equation}
where $m$ is the azimuthal number of the mode. It has the form of the eigenvalue problem 
$(\mathcal{L}-\Lambda)R=0$ with $\mathcal{M}=- \frac{1}{\rho} \frac{\ddroit }{\ddroit \rho}\left( {\rho} \frac{\ddroit}{\ddroit \rho}\right)
+\frac{m^2}{\rho^2}$ and $\Lambda=\varepsilon\mu k^2$ and the continuous spectrum of the operator $\mathcal{L}$ is the real axis. \\
The transformation to obtain cylindrical PML only acts on the radial variable and is given by $\widetilde{\rho}=s_{\rho}\rho$, with $s=\sigma\e^{\ic\phi}$. 
Substituting $\widetilde{\rho}$ into Eq.~(\ref{eq_Bessel_rad}), we obtain a similar spectral problem $(\mathcal{M}-\widetilde{\Lambda})\widetilde{R}=0$, with 
$\widetilde{\Lambda}=\Lambda/s_\rho^2$. Since $s_\rho$ is complex, one can see that the effect of adding infinite cylindrical PML rotates the real positive 
continuous spectrum in the complex plane of an angle $-\phi$ which is now the half-line with parametric equation 
$$\omega(\Lambda)=\frac{c}{s_\rho}\sqrt{\frac{\Lambda}{\varepsilon\mu}},$$
with $\Lambda\in\mathbb{R}^+$.

\subsection{The monoperiodic case with Cartesian PMLs}

The periodicity along $(Ox)$ impose seeking for solutions $v$ verifying Bloch decomposition:
\begin{equation}
v(x,y)=\sum_{m\in\mathbb{Z}}v_{ym}^d(y)\e^{\ic\alpha_mx}
\end{equation}
with $\alpha_m=\alpha+\frac{2\pi}{d}m$.
Inserting this decomposition in Eq.~(\ref{eq:eigenpb}) reads:
\begin{equation}
 -\frac{\ddroit^2 v_{ym}}{\ddroit y^2}=\Lambda_{m} v_{ym}
 \label{SPgrating}
\end{equation}
with
\begin{equation}
   \Lambda_{m}=\omega^2/c^2\varepsilon\mu-\alpha_m^2.
  \label{eqbetanm}
     \end{equation}

The problem then boils down to the spectral study of the canonical operator $\mathcal{M}=-\frac{\ddroit^2}{\ddroit y^2}$, 
which continuous spectrum is $\mathbb{R}^+$. For the grating problem, the continuous spectrum is 
thus composed of several half-lines on the real axis, corresponding to different diffraction orders in the substrate and the superstrate, 
and given by the parametric equations
\begin{equation}
\omega(\Lambda)=\mathcal{D}_m^\pm(\Lambda)=c^\pm \sqrt{\alpha_m^2+\Lambda},\hspace{10pt} \Lambda\in\mathbb{R}^+, \hspace{10pt} \forall m\in\mathbb{Z}
\label{SC_sansPML}
\end{equation}
with $c^\pm=c/\sqrt{\varepsilon^\pm\mu^\pm}$ the speed of light in the considered medium, \textit{i.e.} the half-lines $[c^\pm\alpha_m,+\infty[$.\\

The Cartesian PMLs used in the periodic case only acts on the $y$ variable, the 
change of coordinates being given by $\widetilde{y}=s_y^\pm y$. Inserting $\widetilde{y}$ in Eq.~(\ref{SPgrating}) leads to 
the family of spectral problems $(\mathcal{M}-\widetilde{\Lambda_m})\widetilde{u_{ym}}=0$, with $\widetilde{\Lambda_m}=\Lambda_m/s_y^\pm$. 
The continuous spectrum of the transformed operator is thus composed of several branches given by the parametric equations
\begin{equation}
\omega(\Lambda)=c^\pm \sqrt{\alpha_m^2+\frac{\Lambda}{{s_y^\pm}^2}},\hspace{10pt} \Lambda\in\mathbb{R}^+, \hspace{10pt} \forall m\in\mathbb{Z}.
\end{equation}

\section{A brief vocabulary of Spectral Analysis}
\label{appendix:spec_anal}
The localization an classification of the spectrum of an operator $\mathcal{M}$ is based on a derived operator, 
the so-called \emph{resolvent operator}. 
Let us consider an operator $\mathcal{M}:H_1 \longrightarrow H_2$, where $H_1$ and $H_2$ are two Hilbert spaces. 
The resolvent operator $\mathcal{R}_\Lambda(\mathcal{M})$ is defined as \cite{fpcf}: 
\begin{equation}
 \mathcal{R}_\Lambda(\mathcal{M})=(\mathcal{M}-\Lambda\mathcal{I})^{-1}
\end{equation}
where  $\mathcal{I}$ is the identity operator. 
The resolvent set we denote $\rho(\mathcal{M})$ is the set of complex numbers which satisfy the following condition:
\begin{enumerate}
\item  $\mathcal{R}_\Lambda(\mathcal{M})$ exists,
\item  $\mathcal{R}_\Lambda(\mathcal{M})$ is bounded,
\item  $\mathcal{R}_\Lambda(\mathcal{M})$ is dense in $H_2$.
\end{enumerate}
\begin{itemize}
 \item If condition 1 is not fulfilled, we say that $\Lambda$ is an eigenvalue of $\mathcal{M}$ 
 or that $\Lambda$ forms the \emph{point spectrum} of $\mathcal{M}$ which we denote $\sigma_p(\mathcal{M})$.
  \item If condition 1 and 3 but not condition 2 are fulfilled, we say that $\Lambda$ forms the 
  \emph{continuous spectrum} of $\mathcal{M}$ which we denote $\sigma_c(\mathcal{M})$.
    \item If condition 1 and 2 but not condition 3 are fulfilled, we say that $\Lambda$ forms the 
  \emph{residual spectrum} of $\mathcal{M}$ which we denote $\sigma_r(\mathcal{M})$.
\end{itemize}

The \emph{total spectrum} $\sigma(\mathcal{M})$ is the complementary in $\mathbb{C}$ of the resolvent set, we then have:
\begin{equation}
 \sigma(\mathcal{M})=\mathbb{C}\backslash \rho(\mathcal{M})= \sigma_p(\mathcal{M})\cup \sigma_c(\mathcal{M}) \cup \sigma_r(\mathcal{M}).
\end{equation}

In problems generally encountered in electromagnetism as those studied here, it can be shown that the residual spectrum is in fact reduced to the empty set. 
Moreover, the \emph{essential spectrum} we denote $\sigma_{e}(\mathcal{M})$ consists of all points of the spectrum 
except isolated eigenvalues of finite multiplicity. In the cases studied this paper, the point spectrum is the set of isolated eigenvalues of finite multiplicity, 
the essential spectrum and the continuous spectrum can thus be taken to be identical.

\section{Some properties of the adjoint spectral problem}
\label{appendix:prop_adj}
We derive here the expression of the adjoint operator $\Mop^\dagger_{\tens{\xi}}$. By projecting Eq.~(\ref{eq:eigenpb})
on $w$ (we drop hereafter the index $n$) and integrating by parts twice, we obtain:

\begin{eqnarray*}
  \bra \Mop_{\tens{\xi}}(v) \mb  w\ket &=&
-\int_{\Omega}\div(\tens{\xi}\,\grad v)\,\conj{w}\;\ddroit\B r  \\
  &=&\underbrace{-\int_{\Omega}v \,\div(\tens{\xi}\,\grad \conj{w})\,
 \ddroit\B r}_{=\bra  v  \mb   \Mop_{\tens{\xi}^\star}(w)\ket } \\
&&+\underbrace{\int_{\partial{\Omega}}\tens{\xi}\,(v\, \grad \conj{w} -\conj{w}\,\grad v )\cdotp \B n\;\ddroit S}_{=\mathcal{N}_{\tens{\xi}}(v,w)},
\end{eqnarray*}
The first term in the right hand side of the above equality is equal to $\bra  v  \mb  \Mop_{\tens{\xi}^\star}(w)\ket $. 
The second term denoted $\mathcal{N}_{\tens{\xi}}(v,w)$ is a surface term called the \emph{conjunct} \cite{hanson2002operator}. By a suitable choice of the boundary conditions on $\partial{\Omega}$, 
the conjunct vanishes and from this we have $\Mop^\dagger_{\tens{\xi}}=\Mop_{\tens{\xi}^\star}$. The boundary conditions employed is our models are:

\begin{itemize}
 \item Dirichlet homogeneous boundary condition: $v=0$ and $w=0$, which makes the conjunct zero on these boundaries.
 \item Neumann homogeneous boundary condition: $(\tens{\xi} \,\grad v)\cdotp\B n= 0$ and
 $(\tens{\xi}^\star \,\grad w)\cdotp\B n= 0$, which leads to $N_{\tens{\xi}}(v,w)=0$.
  \item Bloch-Floquet quasi-periodicity conditions: let $\Gamma_l$ and $\Gamma_r$ be the two parallels boundaries where to apply these conditions, 
 and $\alpha$ the quasi-periodicity coefficient (a real fixed parameter of the spectral problem). Since $v$ and $w$ are quasiperiodic functions, 
 they can be expressed as $v(x,y)=v_\sharp(y)\e^{\ic\alpha x}$ and $w(x,y)=w_\sharp(y)\e^{\ic\alpha x}$, where $v_\sharp$ and $w_\sharp$ 
 are $d$\nobreakdash-periodic along $x$. We obtain for the conjunct

   \begin{eqnarray*}
&&\int_{\Gamma_r\cup\Gamma_l}\tens{\xi}\,\left[v\, \grad \conj{w} -\conj{w}\,\grad v \right]\cdotp \B n\;\ddroit S\\
&=&\int_{\Gamma_r\cup\Gamma_l}\tens{\xi}\,\left[\,v_\sharp\,\grad \conj{w_\sharp}+\conj{w_\sharp}\,\grad v_\sharp-2\,\ic\,\alpha\, v_\sharp\,\conj{w_\sharp}\,\right] \cdotp \B n\;\ddroit S\\
\end{eqnarray*}

Now since the integrand is~$d$-periodic along $x$, and since the two parallel boundaries are separated by $d$ 
and have normals with opposite directions, the contribution of $\Gamma_r$ and $\Gamma_l$ 
have the same absolute values but are opposite in signs. It means that in the framework of quasiperiodicity, 
the conjunct vanishes too. 
\end{itemize}

We finally get
\begin{eqnarray*}
&&\bra \Mop_{\tens{\xi}}(v) \mb  w\ket =\bra  v \mb \Mop_{\tens{\xi}^\star}(w)\ket \\
&\Leftrightarrow&\bra \Lambda \chi v \mb  w\ket =\bra  v  \mb  \conj{\Lambda} \chi^\dagger w\ket \\
&\Leftrightarrow&\Lambda \bra \chi v  \mb  w\ket ={\Lambda}\bra  \conj{\chi^\dagger} v   \mb w\ket ,
\end{eqnarray*}

which proves that ${\chi^\dagger}=\conj{\chi}$. The adjoint spectral problem takes eventually the form given 
by Eq.~(\ref{eq:eigenpb_adjoint}).\\

We now derive a property of adjoint eigenmodes. Taking the conjugate transpose of Eq.~(\ref{eq:eigenpb}) reads
\begin{equation}
\left\lbrace\Mop_{\tens{\xi}}(\vphantom{\conj{v}}v)\right\rbrace^\star=\Mop_{\tens{\xi}^\star}\left(\conj{v}\right)=\Mop^\dagger_{\tens{\xi}}\left(\conj{v}\right)=\conj{\Lambda}\,\conj{\chi}\,\conj{v}.
\label{conj_adj}
\end{equation}
It is tempting from Eq.~(\ref{conj_adj}) to say that $w=\conj{v}$, but one shall remember the boundary conditions. 
Indeed, if we take the conjugate transpose of boundary conditions on $\partial{\Omega}$ for the spectral problem we have:

\begin{itemize}
 \item $\conj {v}=0$ for Dirichlet homogeneous boundary condition,
 \item $(\tens{\xi}^\star \cdotp\mathrm{grad}\, \conj{v})\cdotp\B n= 0$ for Neumann homogeneous boundary condition,
  \item $\conj{v}(x,y)=\conj{v_\sharp}(y)\e^{-\ic\alpha x}$ for quasi-periodicity condition.
\end{itemize}
This means that for a problem with either Neumann or Dirichlet homogeneous boundary conditions, we have $w=\conj{v}$. 
For a non periodic scattering problem, one of these conditions is employed on the outward boundaries of PMLs, 
which means that we only have to solve the spectral problem to obtain the entire set of eigenmodes. In contrast, periodic problems 
lack this nice property. Indeed, the dephasing term imply that $\conj{v}\neq w$ except for $\alpha=0$, and in the general 
case we have to solve the two eigenproblems.


\bibliography{biblio_these.bib}

\end{document}